\documentclass[10pt, prd, twocolumn, nofootinbib,preprint,superscriptaddress]{revtex4}
\pdfoutput=1
\usepackage[T1]{fontenc}
\usepackage{amsmath,amssymb}
\usepackage{epsfig}
\usepackage{float}
\usepackage{graphicx}
\usepackage[usenames,dvipsnames]{color}
\usepackage{subfigure}
\usepackage{slashed}
\usepackage[colorlinks,citecolor=blue]{hyperref}
\usepackage{pdfpages}
\usepackage{color}

\begin{document}
\title{TeV Scale Resonant Leptogenesis with $L_{\mu}-L_{\tau}$ Gauge Symmetry \\ in the Light of Muon $(g-2)$}

\author{Debasish Borah}
\email{dborah@iitg.ac.in}
\affiliation{Department of Physics, Indian Institute of Technology Guwahati, Assam 781039, India}

\author{Arnab Dasgupta}
\email{arnabdasgupta@protonmail.ch}
\affiliation{Institute of Convergence Fundamental Studies , Seoul-Tech, Seoul 139-743, Korea}

\author{Devabrat Mahanta}
\email{devab176121007@iitg.ac.in}
\affiliation{Department of Physics, Indian Institute of Technology Guwahati, Assam 781039, India}

\begin{abstract}
Motivated by the growing evidence for the possible lepton flavour universality violation after the first results from Fermilab's muon $(g-2)$ measurement, we revisit one of the most widely studied anomaly free extensions of the standard model namely, gauged $L_{\mu}-L_{\tau}$ model, to find a common explanation for muon $(g-2)$ as well as baryon asymmetry of the universe via leptogenesis. The minimal setup allows TeV scale resonant leptogenesis satisfying light neutrino data while the existence of light $L_{\mu}-L_{\tau}$ gauge boson affects the scale of leptogenesis as the right handed neutrinos are charged under it. For $L_{\mu}-L_{\tau}$ gauge boson mass at GeV scale or above, the muon $(g-2)$ favoured parameter space is already ruled out by other experimental data while bringing down its mass to sub-GeV regime leads to vanishing lepton asymmetry due to highly restrictive structures of lepton mass matrices at the scale of leptogenesis. Extending the minimal model with two additional Higgs doublets can lead to a scenario consistent with successful resonant leptogenesis and muon $(g-2)$ while satisfying all relevant experimental data.
\end{abstract}

\maketitle
\noindent
{\bf Introduction}: The recent measurement of the muon anomalous magnetic moment, $a_\mu$ = $(g - 2)_\mu/2$, by the E989
experiment at Fermilab for the first time shows a discrepancy with respect to the theoretical prediction of the Standard
Model (SM) \cite{Abi:2021gix}
\begin{eqnarray}
a^{\rm FNAL}_\mu = 116 592 040(54) \times 10^{-11}\\
a^{\rm SM}_\mu = 116 591 810(43) \times 10^{-11}
\end{eqnarray}
which, when combined with the previous Brookhaven determination of
\begin{equation}
a^{\rm BNL}_\mu = 116 592 089(63) \times 10^{-11}
\end{equation}
leads to a 4.2 $\sigma$ observed excess of
$\Delta a_\mu = 251(59) \times 10^{-11}$ \footnote{The latest lattice results \cite{Borsanyi:2020mff} however, predict a larger value of muon $(g-2)$ bringing it closer to experimental value. Tension of measured muon $(g-2)$ with global electroweak fits from $e^+ e^-$ to hadron data has also been reported in \cite{Crivellin:2020zul}.}. The status of the SM calculation of muon magnetic moment has been updated recently in \cite{Aoyama:2020ynm}. For more details, one may refer to \cite{Zyla:2020zbs}. The latest Fermilab measurements have also led to several recent works on updating possible theoretical models with new data, a comprehensive review of which may be found in \cite{Athron:2021iuf}. Gauged lepton flavour models like $U(1)_{L_{\mu}-L_{\tau}}$ provide a natural origin of muon $(g-2)$ in a very minimal setup while also addressing the question of light neutrino mass simultaneously. Recent studies on this model related to muon $(g-2)$ may be found in \cite{Borah:2020jzi, Zu:2021odn, Amaral:2021rzw, Zhou:2021vnf, Borah:2021jzu}. While this could be due to lepton flavour universality (LFU) violation, similar anomalies, that too in the context of muon, comes from the measurement of $R_K = {\rm BR}(B \rightarrow K \mu^+ \mu^-)/{\rm BR}(B \rightarrow K e^+ e^-)$. While the hint for this anomaly, like muon $(g-2)$ was there for several years, recent update from the LHCb collaboration \cite{Aaij:2021vac} has led to the most precise measurement ever with more than $3\sigma$ deviation from the SM predictions. In the light of growing evidences for such LFU violations, need for beyond standard model physics around the TeV corner has become very prominent.

Here we consider the popular and minimal model based on the gauged $L_{\mu}-L_{\tau}$ symmetry which is anomaly free \cite{He:1990pn, He:1991qd}.  Apart from the SM fermion content, the minimal version of this model has three heavy right handed neutrinos (RHN) leading to to type I seesaw origin of light neutrino masses \cite{Minkowski:1977sc, Mohapatra:1979ia, Yanagida:1979as, GellMann:1980vs, Glashow:1979nm, Schechter:1980gr}. The same RHNs can lead to leptogenesis \cite{Fukugita:1986hr, Davidson:2008bu} via out-of-equilibrium decay into SM leptons. However, with hierarchical RHN spectrum, there exists a lower bound on the scale of leptogenesis, known as the Davidson-Ibarra bound $M_1 > 10^9$ GeV \cite{Davidson:2002qv}. Several earlier works \cite{Adhikary:2006rf, Chun:2007vh, Ota:2006xr, Asai:2017ryy, Asai:2020qax} considered different scenarios like supersymmetry, high scale leptogenesis within type I seesaw framework of this model. However, it is also possible to have TeV scale leptogenesis via resonant enhancement due to tiny mass splitting between RHNs, known as the resonant leptogenesis \cite{Pilaftsis:1998pd, Pilaftsis:2003gt, Moffat:2018wke, Dev:2017wwc}. It should be noted that earlier works on leptogenesis in gauged $L_{\mu}-L_{\tau}$ model considered high scale breaking of such gauge symmetry and compatibility with muon $(g-2)$ explanation from a low scale vector boson was missing. We intend to perform a general analysis as well as to bridge this gap showing the possibility of TeV scale leptogenesis along with muon $(g-2)$ from a light $L_{\mu}-L_{\tau}$ vector boson.

Motivated by the recent measurements of the muon $(g-2)$, in this work we consider the possibility of low scale leptogenesis and constrain the $L_{\mu}-L_{\tau}$ gauge sector from the requirements of successful leptogenesis and $(g-2)_{\mu}$. Since the decaying RHNs can have this new gauge interaction which can keep them in equilibrium for a longer epochs and can also initiate some washout processes, the requirement of successful leptogenesis for a fixed scale of leptogenesis around a TeV can lead to constraints on the gauge sector parameter space. Since leptogenesis is a high scale phenomena and the requirement of $(g-2)_{\mu}$ needs a low scale breaking $L_{\mu}-L_{\tau}$ gauge symmetry, one can not realise both in the minimal version of the model. We first discuss the minimal model from the requirement of satisfying neutrino data and baryon asymmetry from leptogenesis and then consider an extended model which can accommodate muon $(g-2)$ as well. While we do not pursue the study of $R_K$ anomalies in this model, one may refer to \cite{Biswas:2019twf} for common origin of muon $(g-2)$ and $R_K$ anomalies along with dark matter in extensions of minimal $L_{\mu}-L_{\tau}$ model. Explanation of similar flavour anomalies along with muon $(g-2)$ in this model have also been studied \cite{Crivellin:2015mga, Altmannshofer:2016oaq}. Recently, a dark matter extension of the $L_{\mu}-L_{\tau}$ model was also found to provide a common origin of muon $(g-2)$ and electron recoil excess reported by the XENON1T collaboration \cite{Borah:2020jzi, Borah:2021jzu}.
 
\begin{table}[h!]
\small
	\begin{center}
		\begin{tabular}{||@{\hspace{0cm}}c@{\hspace{0cm}}|@{\hspace{0cm}}c@{\hspace{0cm}}|@{\hspace{0cm}}c@{\hspace{0cm}}|@{\hspace{0cm}}c@{\hspace{0cm}}||}
			\hline
			\hline
			\begin{tabular}{c}
				{\bf ~~~~ Gauge~~~~}\\
				{\bf ~~~~Group~~~~}\\ 
				\hline
				
				$SU(2)_{L}$\\ 
				\hline
				$U(1)_{Y}$\\ 
				\hline
				$U(1)_{L_\mu-L_\tau}$\\ 
			\end{tabular}
			&
			&
			\begin{tabular}{c|c|c}
				\multicolumn{3}{c}{\bf Fermion Fields}\\
				\hline
				~~~$N_e$~~~& ~~~$N_{\mu}$~~~ & ~~~$N_{\tau}$~~~ \\
				\hline
				$1$&$1$&$1$\\
				\hline
				$0$&$0$&$0$\\
				\hline
				$0$&$1$&$-1$\\
			\end{tabular}
			&
			\begin{tabular}{c|c}
				\multicolumn{2}{c}{\bf Scalar Field}\\
				\hline
				~~~$\Phi_{1}$~~~& ~~~$\Phi_2$~~~ \\
				\hline
				$1$&$1$\\
				\hline
				$0$&$0$\\
				\hline
				$1$&$2$\\
			\end{tabular}\\
			\hline
			\hline
		\end{tabular}
		\caption{New Particles and their corresponding
			gauge charges in the minimal model.}
		\label{tab1}
	\end{center}    
\end{table}

\noindent
{\bf Minimal Gauged $L_{\mu}-L_{\tau}$ Model}:
The SM fermion content with their gauge charges under $SU(3)_c \times SU(2)_L \times U(1)_Y \times U(1)_{L_{\mu}-L_{\tau}}$ gauge symmetry are denoted as follows.

$$ q_L=\begin{pmatrix}u_{L}\\
	d_{L}\end{pmatrix} \sim (3, 2, \frac{1}{6}, 0), \; u_R (d_R) \sim (3, 1, \frac{2}{3} (-\frac{1}{3}), 0)$$
$$L_e=\begin{pmatrix}\nu_{e}\\
	e_{L}\end{pmatrix} \sim (1, 2, -\frac{1}{2}, 0), \; e_R \sim (1, 1, -1, 0) $$
$$L_{\mu}=\begin{pmatrix}\nu_{\mu}\\
	\mu_{L}\end{pmatrix} \sim (1, 2, -\frac{1}{2}, 1), \; \mu_R \sim (1, 1, -1, 1) $$
$$L_{\tau}=\begin{pmatrix}\nu_{\tau}\\
	\tau_{L}\end{pmatrix} \sim (1, 2, -\frac{1}{2}, -1), \;  \tau_R \sim (1, 1, -1, -1)$$ 

The new field content apart from the SM ones are shown in table \ref{tab1}.	Only the second and third generations of leptons are charged under the $L_{\mu}-L_{\tau}$ gauge symmetry. The relevant Lagrangian can be written as
\begin{widetext}
\begin{align}
\mathcal{L} & \supseteq \overline{N_{\mu}} i \gamma^\mu D_\mu N_{\mu} - \frac{M_{\mu \tau}}{2} N_{\mu} N_{\tau} +\overline{N_{\tau}} i \gamma^\mu D_\mu N_{\tau} - \frac{M_{ee}}{2} N_e N_e -Y_{e\mu} \Phi^{\dagger}_1 N_e N_\mu -Y_{e\tau} \Phi_1 N_e N_\tau -Y_{\mu} \Phi^{\dagger}_2 N_\mu N_\mu \nonumber \\&-Y_{De} \bar{L_e} \tilde{H} N_e-Y_{D\mu} \bar{L_\mu} \tilde{H} N_\mu-Y_{D \tau} \bar{L_\tau} \tilde{H} N_\tau-Y_{\tau} \Phi_2 N_\tau N_\tau - Y_{le} \overline{L_e} H e_R + Y_{l\mu} \overline{L_\mu} H \mu_R
             + Y_{l\tau} \overline{L_\tau} H \tau_R+ {\rm h.c.}
             \label{yuklag}
\end{align}
\end{widetext}
where $H$ is the SM Higgs doublet.  The covariant derivatives for RHNs are defined by
\begin{eqnarray}
\slashed{D}N_{\mu} &= &(\slashed{\partial}+i x g_{\mu \tau} \slashed{Z}_{\mu \tau}) N_{\mu}, \\
\slashed{D}N_{\tau} &= &(\slashed{\partial}-i x g_{\mu \tau} \slashed{Z}_{\mu \tau}) N_{\tau} .
\end{eqnarray}
While the neutral component of the Higgs doublet $H$ breaks the electroweak gauge symmetry, the singlets break $L_{\mu}-L_{\tau}$ gauge symmetry after acquiring non-zero vacuum expectation values (VEV). Denoting the VEVs of singlets $\Phi_{1,2}$ as $v_{1,2}$, the new gauge boson mass can be found to be $M_{Z_{\mu \tau}}=g_{\mu \tau} \sqrt{(v^2_1+4v^2_2)}$ with $g_{\mu \tau}$ being the $L_{\mu}-L_{\tau}$ gauge coupling. Clearly the model predicts diagonal charged lepton mass matrix $M_\ell$ and diagonal Dirac Yukawa of light neutrinos. Thus, the non-trivial neutrino mixing will arise from the structure of right handed neutrino mass matrix $M_R$ only which is generated by the chosen scalar singlet fields. The right handed neutrino mass matrix, Dirac neutrino mass matrix and charged lepton mass matrix are given by
\begin{widetext}
\begin{align}
M_R =\begin{pmatrix}
               M_{ee}      &  Y_{e\mu} \frac{v_1}{\sqrt{2}}
    & Y_{e\tau}  \frac{v_1}{\sqrt{2}} \\
               Y_{e\mu}  \frac{v_1}{\sqrt{2}}      &  \sqrt{2} Y_{\mu}
 v_2     & \frac{M_{\mu \tau}}{2}  \\
               Y_{e\tau}  \frac{v_1}{\sqrt{2}}     &  \frac{M_{\mu \tau}}{2}    &
               \sqrt{2} Y_{\tau}  v_2 
               \end{pmatrix}\, , \;\;
M_D =\begin{pmatrix}
               Y_{De} \frac{v}{\sqrt{2}}       &  0    & 0  \\
               0     &  Y_{D\mu} \frac{v}{\sqrt{2}}    & 0  \\
               0     &  0    & Y_{D \tau} \frac{v}{\sqrt{2}}  
               \end{pmatrix},\,              M_\ell= \begin{pmatrix}
Y_{le} \frac{v}{\sqrt{2}}  & 0 & 0\\
0 & Y_{l\mu}\frac{v}{\sqrt{2}}  & 0 \\
0 & 0 & Y_{l\tau} \frac{v}{\sqrt{2}} 
\end{pmatrix}
\label{massmatrices}
\end{align}
\end{widetext}
Here $v$ is the VEV of neutral component of SM Higgs doublet $H$. We can find the light neutrino mass matrix can be found by the type I seesaw formula as
\begin{eqnarray}
M_{\nu} & = &-M_{D}M_{R}^{-1}M_{D}^{T} \\
        & = & \begin{pmatrix}
        A & B & C \\ B & D & E \\ 
        C & E & F
        \end{pmatrix}.
\end{eqnarray}
The mass matrices for RHN as well as light neutrinos in this model do not possess any specific structure and hence it is straightforward to fit the neutrino oscillation data. In the presence of only one scalar singlet $\Phi_1$, the RHN mass matrix has zero entries at $(\mu \mu)$ and $(\tau \tau)$ entries leading to a two-zero minor structure \cite{Asai:2017ryy} in $M^{-1}_{\nu}$. The light neutrino mass matrix, although does not contain any zeros, leads to two constraints among its elements. In view of tight constraints on neutrino parameters from global fit data \cite{Esteban:2018azc, deSalas:2020pgw, Zyla:2020zbs} as well as cosmology bounds on sum of light neutrino masses from Planck 2018 data \cite{Aghanim:2018eyx}, it is difficult to satisfy the data using the constrained structure of mass matrices. Similar conclusion was also arrived at in earlier works \cite{Asai:2017ryy, Asai:2020qax}. This becomes more restrictive when we constrain two of the RHNs to be in TeV regime with tiny mass splittings for resonant leptogenesis. Therefore, we have introduced another singlet scalar $\Phi_2$ which gives rise to general structures of $M_R$ and $M_{\nu}$.

It should be noted that, a kinetic mixing term between $U(1)_Y$ of SM and $U(1)_{L_{\mu}-L_{\tau}}$ of the form $\frac{\epsilon}{2} B^{\alpha \beta} Y_{\alpha \beta}$ can exist in the Lagrangian where $B^{\alpha\beta}=  \partial^{\alpha}X^{\beta}-\partial^{\beta}X^{\alpha}, Y_{\alpha \beta}$ are the field strength tensors of $U(1)_{L_{\mu}-L_{\tau}}, U(1)_Y$ respectively and $\epsilon$ is the mixing parameter. Even if this mixing is considered to be absent in the Lagrangian, it can arise at one loop level with particles charged under both the gauge sectors in the loop. We consider this mixing to be $\epsilon = g_{\mu \tau}/70$. While the phenomenology of muon $(g-2)$, leptogenesis in our model is not much dependent on this mixing, the experimental constraints on the model parameters can crucially depend upon this mixing. We therefore choose it to be small, around the same order as its one loop value.

\noindent
{\bf Anomalous Muon Magnetic Moment}: The magnetic moment of muon is defined as
\begin{equation}\label{anomaly}
\overrightarrow{\mu_\mu}= g_\mu \left (\frac{q}{2m} \right)
\overrightarrow{S}\,,
\end{equation}
where $g_\mu$ is the gyromagnetic ratio and its value is $2$ for an elementary spin $\frac{1}{2}$ particle of mass $m$ and charge
$q$. However, higher order radiative corrections can generate additional contributions to its magnetic moment and is parameterised as
\begin{equation}
a_\mu=\frac{1}{2} ( g_\mu - 2).
\end{equation}
As mentioned earlier, the anomalous muon magnetic moment has been measured very precisely in the recent Fermilab experiment while it has also been predicted in the SM to a great accuracy.
In the model under consideration in this work, the additional contribution to muon magnetic moment arises dominantly from one loop diagram mediated by $L_{\mu}-L_{\tau}$ gauge boson $Z_{\mu \tau}$. The corresponding one loop contribution is given by \cite{Brodsky:1967sr, Baek:2008nz}
\begin{equation}
\Delta a_{\mu} = \frac{\alpha'}{2\pi} \int^1_0 dx \frac{2m^2_{\mu} x^2 (1-x)}{x^2 m^2_{\mu}+(1-x)M^2_{Z'}} \approx \frac{\alpha'}{2\pi} \frac{2m^2_{\mu}}{3M^2_{Z'}}
\end{equation}
where $\alpha'=g^2_{\mu \tau}/(4\pi)$. Note that in the presence of additional Higgs doublets (a scenario which we will discuss in the upcoming sections) we can have another one-loop diagram mediated by charged component of such additional scalar doublet and right handed neutrino $N_{\mu}$. However, since such particles in loop couple only to the left handed muons, the required chirality flip has to occur in external muon legs, leading to a suppressed contribution to muon $(g-2)$. While there can be a very fine-tuned parameter space with large Yukawa couplings and specific loop particle masses \cite{Calibbi:2018rzv}, it is unlikely to be in agreement with light neutrino mass data along with successful TeV scale leptogenesis requirements. We therefore do not study this fine-tuned possibility here and focus on the light neutral gauge boson contribution alone.\\

\noindent
{\bf Resonant Leptogenesis}:
In this section we study the possibility of leptogenesis from the out of equilibrium decays of RHNs neutrinos $N_{i}$. The generated $B-L$ asymmetry can be converted into a baryon asymmetry by the sphaleron processes which conserves $B-L$ asymmetry but violet $B+L$ asymmetry. The sphaleron processes are active between temperatures of
$10^{12}$ GeV to  $10^{2}$ GeV in the early Universe. At high temperatures the sphalerons are in
thermal equilibrium and subsequently they freeze-out just before the electroweak symmetry breaking (EWSB) at around $100 \; {\rm GeV} < T < 200\; {\rm GeV}$. Usually the mass of required to generated the observe asymmetry lies above   the scale of $10^{8}-10^{9}$ GeV in conventional vanilla leptogenesis \cite{Plumacher:1996kc,Buchmuller:2002rq}. However, the detection of these very heavy right handed neutrinos is beyond
the reach of LHC and other near future colliders. That is why we are mainly focusing on TeV scale leptogenesis in this work by exploiting the resonance enhancement condition ($M_{2}-M_{1}\sim \Gamma_{1}/2$). This framework is known as resonant leptogenesis  \cite{Pilaftsis:1997jf,Pilaftsis:2003gt,Heeck:2016oda}. The relevant Yukawa matrix for leptogenesis can be identified to be 

\begin{equation}
h=\begin{pmatrix}
Y_{De}V_{11} & Y_{De}V_{12} & Y_{De}V_{13} \\
Y_{D \mu}V_{21} & Y_{D\mu}V_{22} & Y_{D\mu}V_{23} \\
Y_{D\tau}V_{31} & Y_{D\tau}V_{32} & Y_{D\tau} V_{33},
\end{pmatrix} \label{eq:Yukawa}
\end{equation}
where $V_{ij}$ are the elements of the matrix $V$ which diagonalises $M_{R}$.
\begin{equation}
{\rm diag}(M_{1},M_{2},M_{3})=V^{\dagger} M_{R}V^*
\end{equation}

The CP asymmetry parameter corresponding to the CP violating decay of RHN $N_i$ (summing over all lepton flavours) is given by \cite{Pilaftsis:2003gt}
\begin{eqnarray}
\epsilon_{i} & = & \dfrac{\Gamma_{(N_{i}\longrightarrow \sum_{\alpha} L_{\alpha} H )}-\Gamma_{(N_{i}\longrightarrow \sum_{\alpha} L_{\alpha}^{c}H)}}{\Gamma_{(N_{i}\longrightarrow \sum_{\alpha}L_{\alpha}H)}+\Gamma_{(N_{i}\longrightarrow \sum_{i}L_{\alpha}^{c}H)}}  \\ 
 & = &\dfrac{{\rm Im}[(h^{\dagger}h)_{ij}^{2}]}{(h^{\dagger}h)_{ii}(h^{\dagger}h)_{jj}}\dfrac{(M_{i}^{2}-M_{j}^{2})M_{i}\Gamma_{j}}{(M_{i}^{2}-M_{j}^{2})^{2}+M_{i}^{2}\Gamma_{j}^{2}}. \label{eq:asymmparameter}
\end{eqnarray}

Since we are mainly focusing on the parameter space where $M_{1} \backsimeq M_{2}<M_{3}$ such that the leptogenesis is mainly governed by the resonant enhancement between $N_{1}$ and $N_{2}$ and the CP asymmetry coming from the decay of $N_{3}$ is negligible. The resonance condition is satisfied when $M_2-M_1 \simeq \Gamma_1/2$ where $\Gamma_1$ is the decay width of the lightest RHN $N_1$. The relevant asymmetry parameters namely, $\epsilon_{1}$ and $\epsilon_{2}$ crucially depend upon $Im[(h^{\dagger}h)^{2}_{12}]$ and $Im[(h^{\dagger}h)^{2}_{21}]$ respectively. From equation \eqref{eq:Yukawa} one can write 

\begin{eqnarray}
{\rm Im}[(h^{\dagger}h)_{12}^{2}]& = & -{\rm Im}[(h^{\dagger}h)_{21}^{2}] = \dfrac{1}{2}{\rm Im}[( \mid Y_{De} \mid^{2} V_{12}V_{11}^{*}+\nonumber \\   && \mid Y_{D \mu} \mid^{2}V_{22}V_{21}^{*}
  + \mid Y_{D \tau} \mid^{2} V_{32} V_{31}^{*} )^{2}] \nonumber \\ \label{eq:complex1}
\end{eqnarray}
From equation \eqref{eq:complex1} the importance of the phases appearing in the elements of the matrix $V$ can be seen. These phases are appearing in the diagonalising matrix $V$ from the complex parameters in $M_{R}$. 

The relevant Boltzmann equations for our setup can be written as 

\begin{eqnarray}
\dfrac{dn_{N_{1}}}{dz} & = & D_{1}(n_{N_{1}}-n_{N_{1}}^{\rm eq}) - \dfrac{s}{{\bf H}z}  \left( (n_{N_{1}})^{2}-(n_{N_{1}}^{\rm eq})^{2}\right) \nonumber \\  & & \langle \sigma v\rangle_{\small{N_{1}N_{1}\longrightarrow X X}},  \\
\dfrac{dn_{N_{2}}}{dz} & = & D_{2}(n_{N_{2}}-n_{N_{2}}^{\rm eq}) - \dfrac{s}{{\bf H}z}  \left( (n_{N_{2}})^{2}-(n_{N_{2}}^{\rm eq})^{2}\right) \nonumber \\  & & \langle \sigma v \rangle_{\small{N_{2}N_{2}\longrightarrow X X}}, \\
\dfrac{dn_{B-L}}{dz} &=& -\epsilon_{1}D_{1}\left( n_{N_{1}}-n_{N_{1}}^{\rm eq} \right)-\epsilon_{2}D_{2}\left( n_{N_{2}}-n_{N_{2}}^{\rm eq} \right) \nonumber \\ & & -(WID_{1}+WID_{2}+\Delta W)n_{B-L}, \label{eq:B-L}
\end{eqnarray}
where $n_{N_{i}}$ and $n_{B-L}$ denote the comoving number densities of $N_{i}$ and $B-L$ respectievly. The equilibrium no densities of $N_{i}$'s are defined by $n_{N_{i}}^{\rm eq}=\frac{z^2}{2}\kappa_2(z)$, (with $\kappa_i(z)$ being the modified Bessel function of $i$-th kind) and $z=M_1/T=M_{2}/T$ (for resonant leptogenesis). In $\langle \sigma v \rangle_{N_i N_i \rightarrow X X}$, $X$ denotes any final state particle to which $N_i$'s can annihilate into. $D_{1,2}$ are the decay terms and $WID_{1,2}$ are the inverse decay terms for $N_{1}$ and $N_{2}$ decays respectively, which are measures of the rate of decay and inverse decay with respect to the background expansion of the universe parametrised by Hubble expansion rate $\bf H$. They are defined as

\begin{eqnarray}
D_{i} & = & \dfrac{ \langle \Gamma_{i} \rangle}{{\bf H} z} = K_{i}z\dfrac{\kappa_{1}(z)}{\kappa_{2}(z)},  \\
WID_{i} & = & \dfrac{1}{4}K_{i}z^{3}\kappa_{1}(z),
\end{eqnarray}
with $K_i=\Gamma_i/{\bf H}(z=1)$ being the decay parameter. The term $\Delta W$ on the right hand side of the equation \eqref{eq:B-L} takes account of all the scattering processes that can act as possible washouts for the generated $B-L$ asymmetry. We identify the following scattering washouts in our model, $l W^{\pm} (Z)\longrightarrow N_{1,2} H$, $l Z_{\mu \tau} \longrightarrow N_{1}H $, $q l \longrightarrow q N_{1,2}$, $l N_{1,2} \longrightarrow q q^{c}$, $l H \longrightarrow l^{c}H^*$, $l H \longrightarrow N_{1,2} W^{\pm} (Z)$ and $l N_{1,2}\longrightarrow Z_{\mu \tau},H $. We take all of them into account in our numerical analysis and defined $\Delta W$ as, 

\begin{equation}
\Delta W=\dfrac{\langle \Gamma_{\rm scatterings} \rangle}{{\bf H} z}.
\end{equation}

\begin{figure*}
\includegraphics[scale=.5]{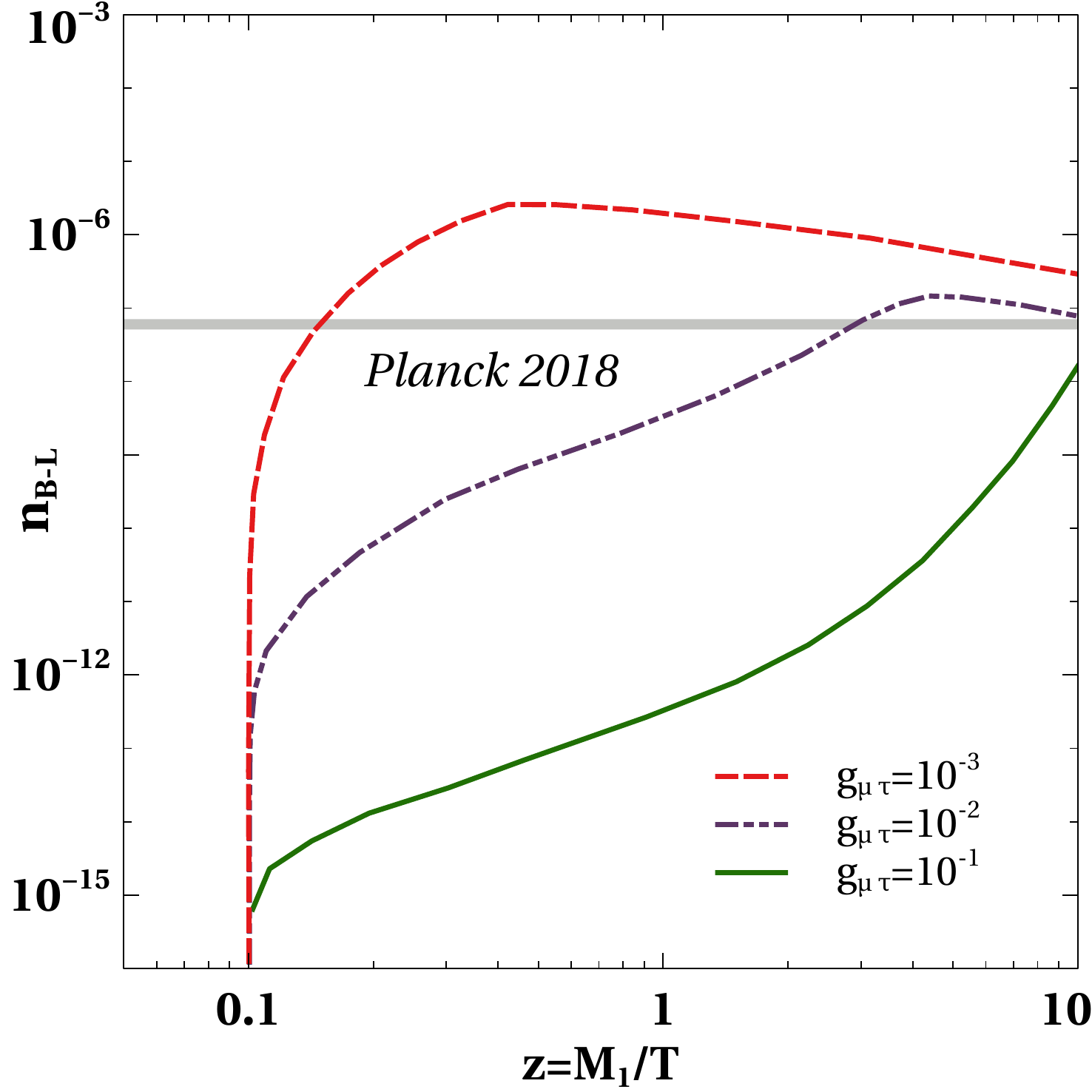}
\includegraphics[scale=.5]{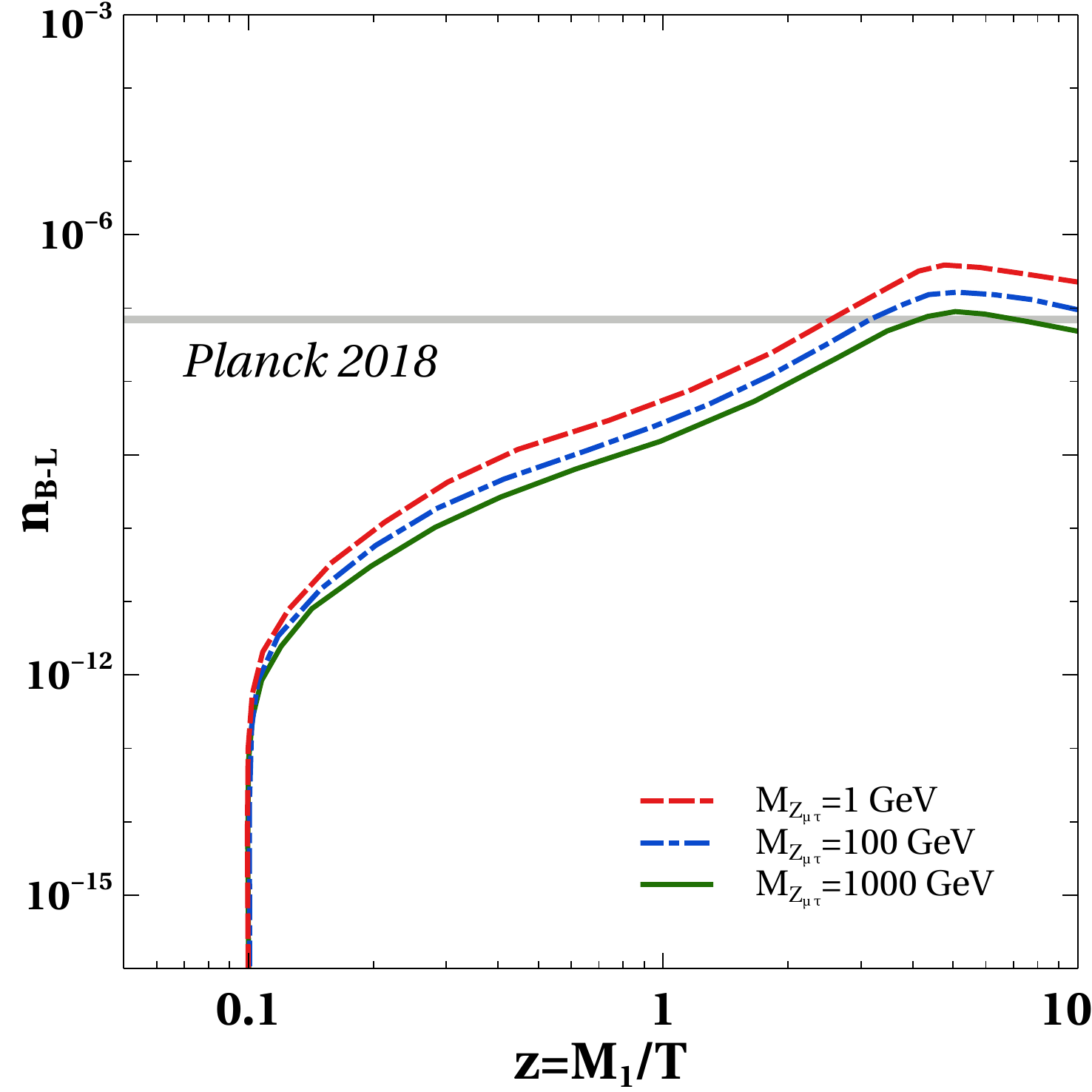}

\caption{Evolution plot of comoving number density of $B-L$ for different benchmark values of $g_{\mu \tau}$ (left panel) and $M_{Z_{\mu \tau}}$ (right panel). The other parameters are set to be $M_{1}=1.38$ TeV, $\Delta M=0.001$ keV, and $M_{Z_{\mu \tau}}=100$ GeV (left panel) and $g_{\mu \tau}=10^{-2}$ (right panel).}
\label{fig:evolution3}
\end{figure*}

\begin{figure*}
 \centering
 \includegraphics[scale=0.48]{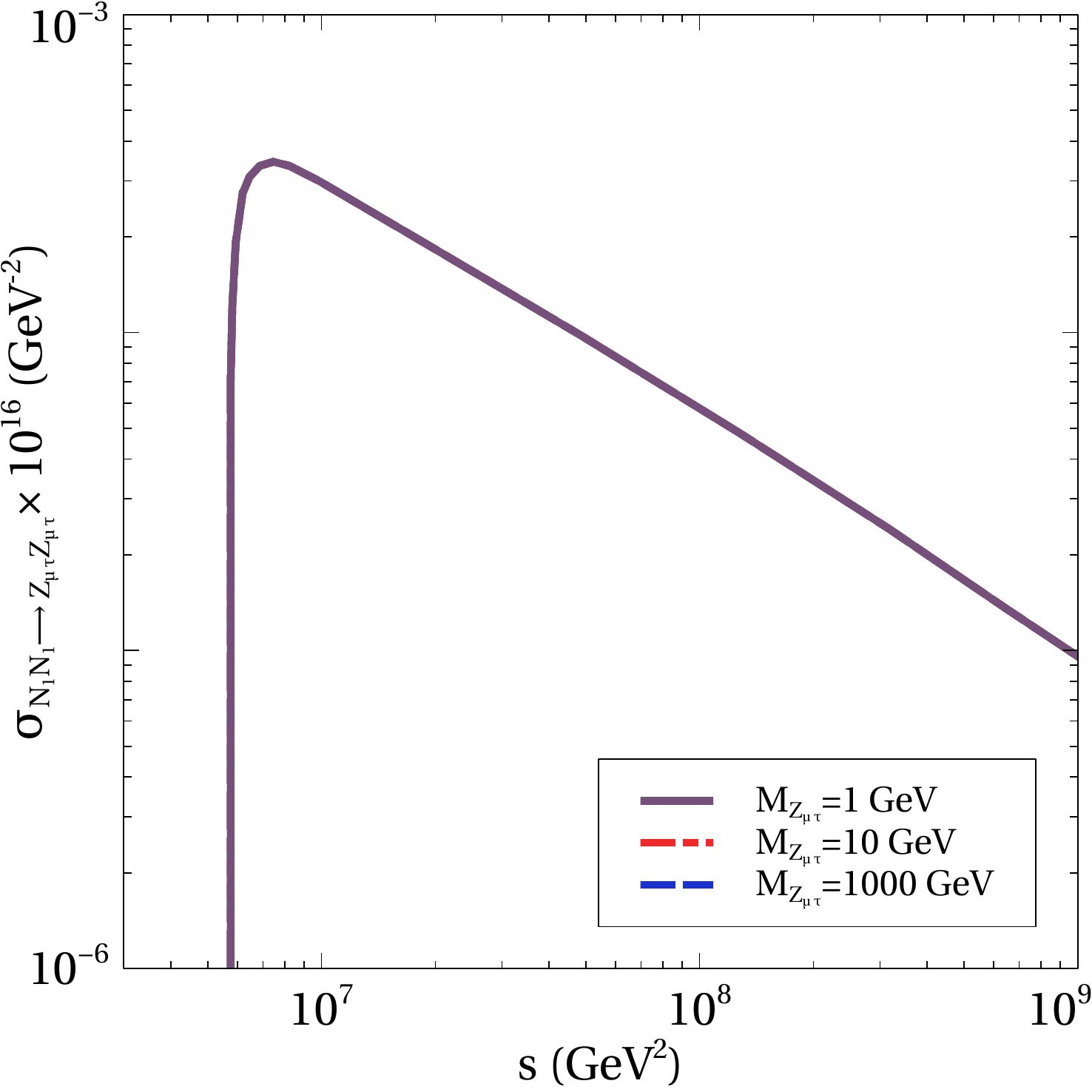}
 \includegraphics[scale=0.48]{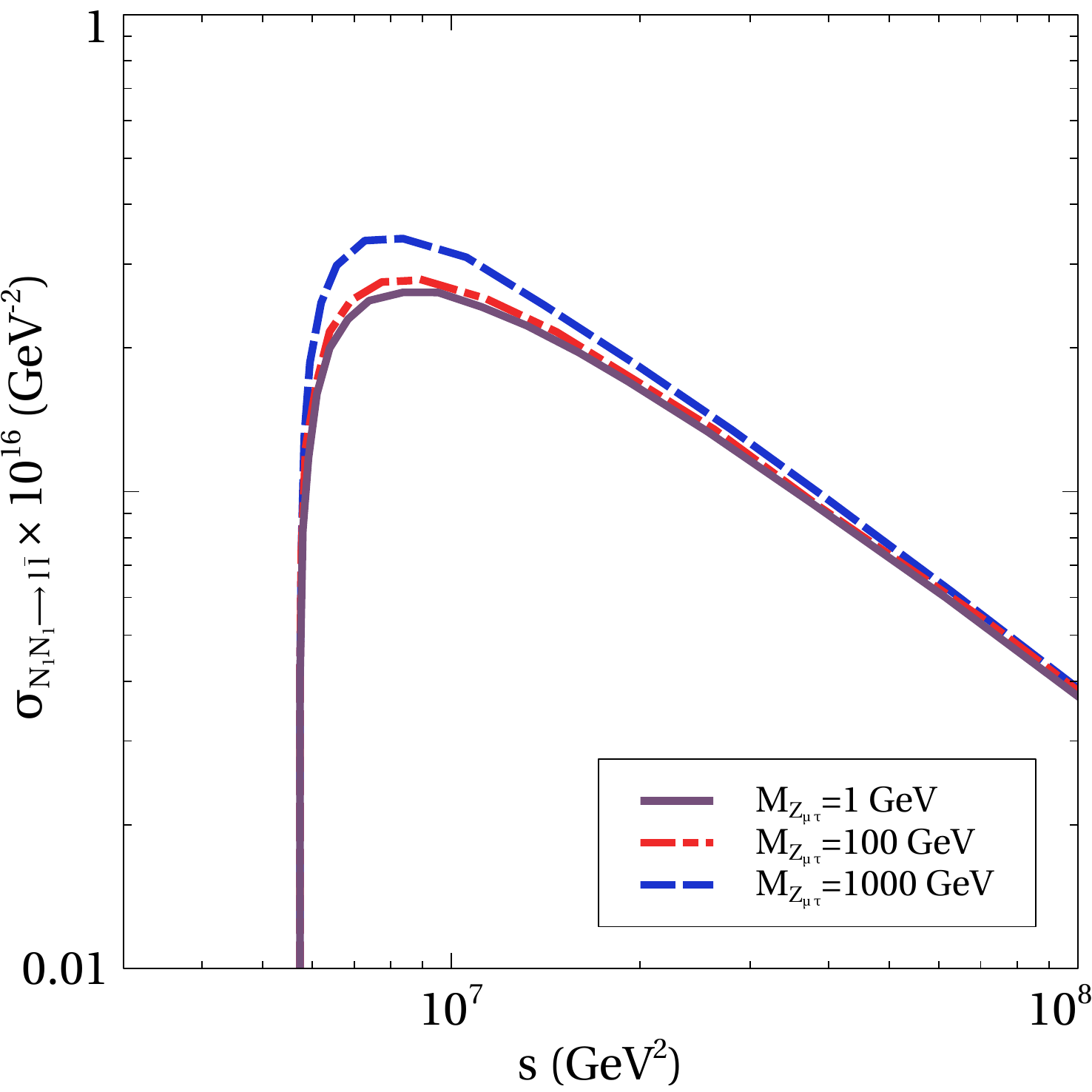}
 \caption{Variation of annihilation cross sections for the processes $N_{1}N_{1} \longrightarrow Z_{\mu \tau} Z_{\mu \tau}$ (left panel), $N_{1}N_{1}\longrightarrow l \bar{l}$ (right panel) with the center of mass energy squared $s$. The relevant benchmark parameters are taken to be $g_{\mu \tau}=0.1$, $M_{1}=1.2$ TeV, $Y_{De} \sim 10^{-8}$, $Y_{D\mu}\sim 10^{-6}$ and $Y_{D\tau}\sim 10^{-6}$. }
 \label{annihilation1}
\end{figure*}  

\begin{figure*}
\includegraphics[scale=.5]{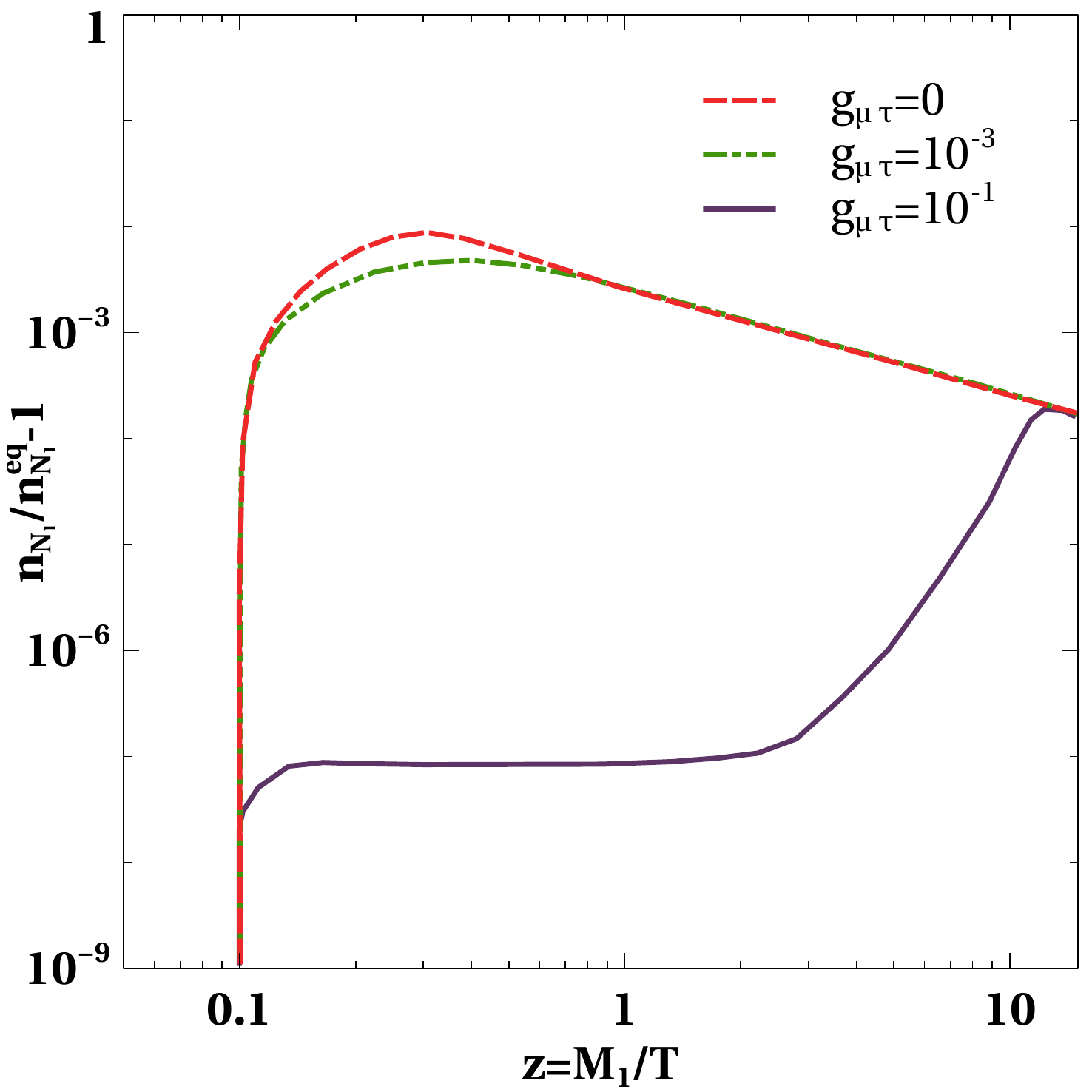}
\includegraphics[scale=.5]{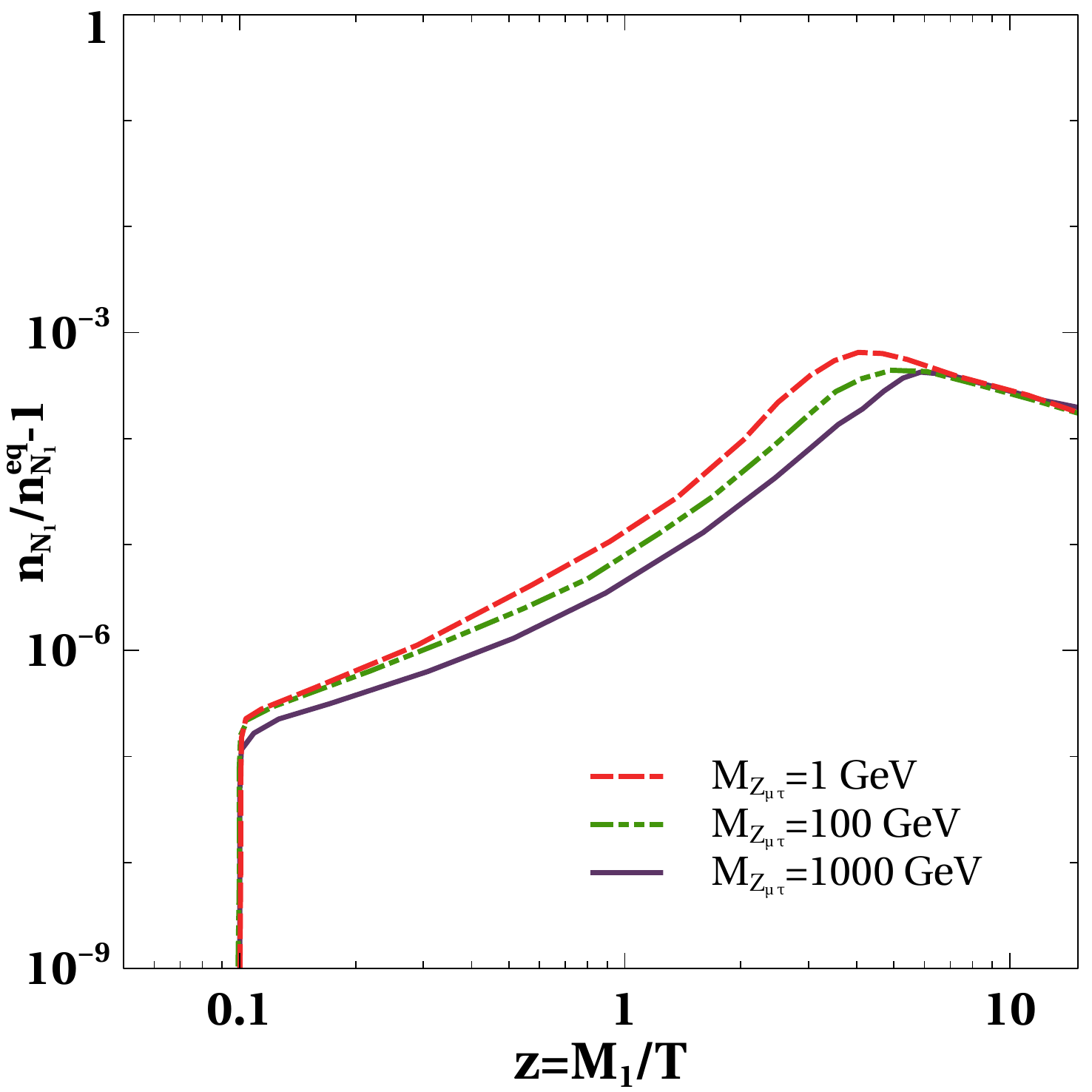}
\caption{Evolution plot of comoving number density of $N_1$ for different benchmark values of $g_{\mu \tau}$ (left panel) and $M_{Z_{\mu \tau}}$ (right panel). The other parameters are set to be $M_{1}=1.38$ TeV, $\Delta M=0.001$ keV and $M_{Z_{\mu \tau}}=200$ GeV (left panel) and $g_{\mu \tau}=10^{-2}$ (right panel).}
\label{fig:evolution3b}
\end{figure*}

Thus, apart from the usual Yukawa or SM gauge coupling related processes, we also have washout processes involving $Z_{\mu \tau}$ gauge boson. Since interactions involving $Z_{\mu \tau}$ can cause dilution of $N_{1}$ abundance as well as wash out the generated lepton asymmetry, one can tightly constrain the $L_{\mu}-L_{\tau}$ gauge sector couplings from the requirement of successful leptogenesis at low scale. Similar discussions on impact of such Abelian gauge sector on leptogenesis can be found in \cite{Iso:2010mv, Okada:2012fs, Heeck:2016oda, Dev:2017xry, Mahanta:2021plx, FileviezPerez:2021ycy}.

\begin{figure}
\centering
\includegraphics[scale=0.5]{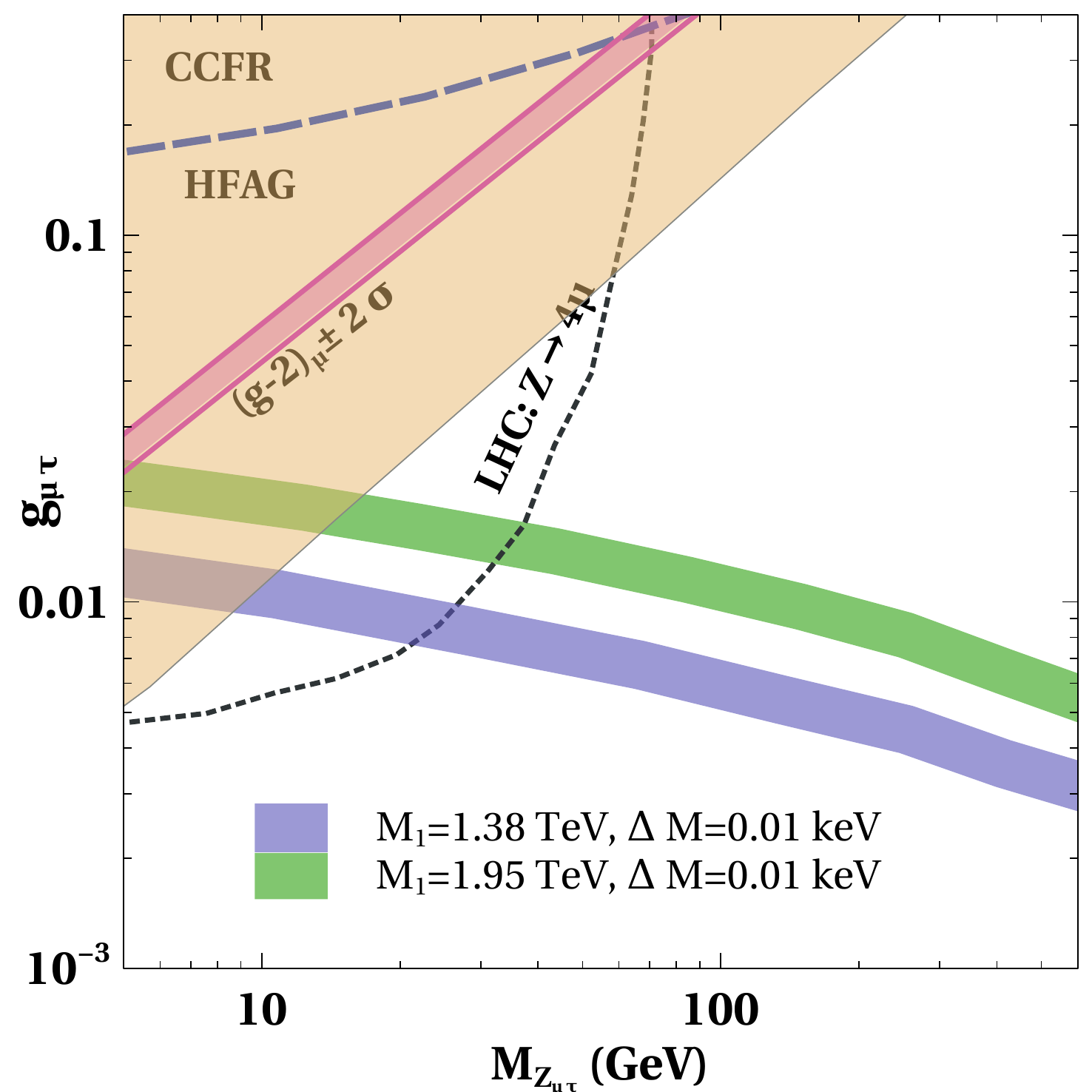}
\caption{Summary plot showing allowed region of parameter space for $M_{Z_{\mu \tau}}$ above GeV. The blue and green coloured bands correspond to the regions where correct baryon asymmetry can be generated for fixed scales of leptogenesis. The pink band represents muon $(g-2)$ favoured region, already ruled out by CCFR exclusion limit \cite{Altmannshofer:2014pba} shown in yellow coloured region. Dashed lines correspond to limits from the LHC \cite{CMS:2018gao} and HFAG lepton universality test \cite{HFLAV:2019otj} (see text for details).}
\label{Fig1}
\end{figure}

After solving the above Boltzmann equations, we convert the final $B-L$ asymmetry $n_{B-L}^f$ just before electroweak sphaleron freeze-out into the observed baryon to photon ratio by the standard formula 
\begin{align}
\eta_B \ = \ \frac{3}{4}\frac{g_*^{0}}{g_*}a_{\rm sph}n_{B-L}^f \ \simeq \ 9.2\times 10^{-3}\: n_{B-L}^f \, ,
\label{eq:etaB}
\end{align}
where $a_{\rm sph}=\frac{8}{23}$ is the sphaleron conversion factor (taking into account two Higgs doublets). Here $g_{*}, g_{*}^{0}$ denote the relativistic degrees of freedom at scale of leptogenesis $T=M_1$ and scale of recombination respectively. The final baryon to photon ratio is then compared with Planck 2018 data \cite{Zyla:2020zbs, Aghanim:2018eyx}
\begin{equation}
\eta_B = \frac{n_{B}-n_{\bar{B}}}{n_{\gamma}} = 6.1 \times 10^{-10}.
\label{etaBobs}
\end{equation}
and constraints on the model parameters are obtained. In figure \ref{fig:evolution3}, we show the evolution of lepton asymmetries for different benchmark values of gauge coupling $g_{\mu \tau}$ (left panel) and gauge boson mass $M_{Z_{\mu \tau}}$ (right panel) while keeping other parameters fixed. We can see that the asymmetry decreases with the increase in $g_{\mu \tau}$. It is because with the increase in $g_{\mu \tau}$ the annihilations of $N_{1,2}$ through $Z_{\mu \tau}$ and to a pair of $Z_{\mu \tau}$ increases, which tries to bring the $N_{1,2}$ number densities close to their equilibrium densities. Thus, depletion in number densities of $N_{1,2}$ leads to a decrease in the final asymmetry.

Again, from figure \ref{fig:evolution3}, we can see that the asymmetry decreases with increase in $M_{Z_{\mu \tau}}$. This is because we have two types of annihilation of $N_{1,2}$ involving $M_{Z_{\mu \tau}}$, the first one being the annihilation into a pair of leptons and the second one is the annihilation into a pair of $M_{Z_{\mu \tau}}$. Out of these two processes, the dominant one is the annihilation into a pair of leptons as can be seen in figure \ref{annihilation1} where the two annihilation cross-sections are shown as a function of center of mass energy squared $s$ for comparison. The annihilation of $N_{1}$ into a pair of lepton is possible through a s-channel diagram mediated by $Z_{\mu \tau}$ and through a t-channel diagram mediated by scalar doublet. The s-channel diagram mediated by singlet scalar due to mixing with the SM like Higgs is ignored as such mixing is generated only after electroweak symmetry breaking. As we are working in the region of parameter space where $M_{Z_{\mu \tau}}<2M_{1}$, therefore with increase in $M_{Z_{\mu \tau}}$ the cross sections for the $Z_{\mu \tau}$ mediated processes increase as seen from the right panel plot of figure \ref{annihilation1}. This tends to bring the $N_{1,2}$ number density closer to its equilibrium density and therefore leading to a decrease in $B-L$ asymmetry.  We show this effect in figure \ref{fig:evolution3b}. Clearly, increase in gauge coupling takes the number density of $N_1$ closer to its equilibrium number density, as seen from the left panel plot of figure \ref{fig:evolution3b} . Increase in $M_{Z_{\mu \tau}}$ also has a similar but milder effect, as seen from the right panel plot of figure \ref{fig:evolution3b}.

In figure \ref{Fig1}, we summarise the results in $g_{\mu \tau}-M_{Z_{\mu \tau}}$ parameter space considering $Z_{\mu \tau}$ mass larger than a GeV. The pink coloured band is the favoured region in $g_{\mu \tau}-M_{Z_{\mu \tau}}$ plane from the Fermilab's result on muon (g-2). The yellow coloured region is excluded by the upper bound on neutrino trident process measured by the CCFR collaboration \cite{Altmannshofer:2014pba}. Therefore one can clearly see that the muon $(g-2)$ favoured region in the high mass regime of $Z_{\mu \tau}$ is completely excluded from the bound on neutrino trident process. Additionally, the LHC bounds searches for multi-lepton final states signatures also rule out some part of the parameter space . We show the exclusion region from LHC measurements of $Z \rightarrow 4 \mu$ \cite{CMS:2018gao} in terms of the dotted line. The region below the blue dashed line is allowed by HFAG lepton universality test at $2\sigma$ level \cite{HFLAV:2019otj}. The leptogenesis favoured parameter space for two different scales of leptogenesis $M_1=1.38$ TeV, $M_1=1.95$ TeV are shown in blue and green coloured bands respectively. As $M_1$ increases, the annihilation rates of $N_1$ decreases, allowing slightly larger values of $g_{\mu \tau}$ for fixed $Z_{\mu \tau}$ mass. On the other hand, as $Z_{\mu \tau}$ mass is increased towards $2M_1$, the annihilation rate of $N_1$ mediated by $Z_{\mu \tau}$ increases, requiring smaller values of $g_{\mu \tau}$ in order to take $N_1$ out of equilibrium resulting in generation of lepton asymmetry.

If sub-GeV regime of $Z_{\mu \tau}$ is explored, there exists an allowed region in $g_{\mu \tau}-M_{Z_{\mu \tau}}$ parameter space consistent with muon $(g-2)$ and other experimental bounds as discussed recently in \cite{Borah:2020jzi, Zu:2021odn, Amaral:2021rzw, Zhou:2021vnf, Borah:2021jzu}. As we will see below, this allowed region correspond to $M_{Z_{\mu \tau}} \sim 10-100$ MeV, $g_{\mu \tau} \sim 10^{-4}-10^{-3}$. However, such sub-GeV $M_{Z_{\mu \tau}}$ will correspond to low scale breaking of $L_{\mu}-L_{\tau}$ gauge symmetry around 100 GeV, even below the sphaleron transition temperature. Prior to the symmetry breaking scale, the RHN mass matrix has very restrictive structure as can be realised by considering vanishing singlet scalar VEVs in $M_R$ given in \eqref{massmatrices}. The fact that the model is inconsistent with successful leptogenesis in the $\mu-\tau$ symmetric limit was also noted in earlier works \cite{Adhikary:2006rf}. Thus, one needs to go beyond the minimal setup in order to find a consistent picture accommodating both leptogenesis and muon $(g-2)$ while agreeing with experimental data including light neutrino mass and mixing. \\

\noindent
{\bf Non-minimal Gauged $L_{\mu}-L_{\tau}$ Model}:

\begin{table}[h!]
	\begin{center}
		\footnotesize
		\begin{tabular}{||@{\hspace{0cm}}c@{\hspace{0cm}}|@{\hspace{0cm}}c@{\hspace{0cm}}|@{\hspace{0cm}}c@{\hspace{0cm}}|@{\hspace{0cm}}c@{\hspace{0cm}}||}
			\hline
			\hline
			\begin{tabular}{c}
				{\bf ~~~~ Gauge~~~~}\\
				{\bf ~~~~Group~~~~}\\ 
				\hline
				
				$\tiny{SU(2)_{L}}$\\ 
				\hline
				$U(1)_{Y}$\\ 
				\hline
				$U(1)_{L_\mu-L_\tau}$\\ 
			\end{tabular}
			&
			&
			\begin{tabular}{c|c|c}
				\multicolumn{3}{c}{\bf Fermion Fields}\\
				\hline
				~~~$N_e$~~~& ~~~$N_{\mu}$~~~ & ~~~$N_{\tau}$~~~ \\
				\hline
				$1$&$1$&$1$\\
				\hline
				$0$&$0$&$0$\\
				\hline
				$0$&$x$&$-x$\\
			\end{tabular}
			&
			\begin{tabular}{c|c|c}
				\multicolumn{3}{c}{\bf Scalar Field}\\
				\hline
				 ~~~$H_2$~~~ & ~~~$H_3$~~~ & ~~~$\phi_{1,2}$~~~  \\
				\hline
				$2$&$2$&$1$\\
				\hline
				$1/2$&$1/2$&$0$\\
				\hline
				$1-x$&$-1+x$&$x,2x$\\
			\end{tabular}\\
			\hline
			\hline
		\end{tabular}
		\caption{New Particles and their corresponding
			gauge charges in the non-minimal model.}
		\label{tab2}
	\end{center}    
\end{table}
In order to achieve successful leptogenesis with sub-GeV $Z_{\mu \tau}$, we extend the minimal model with two additional Higgs doublets as shown in table \ref{tab2}. The presence of these additional Higgs doublets $H_{2, 3}$ are required to allow corresponding Dirac Yukawa couplings with $N_{\mu, \tau}$ whose $L_{\mu}-L_{\tau}$ charges are chosen as $\pm x$ with $x < 1$. While the Higgs doublet $H_1$ is neutral under this additional gauge symmetry, $H_{2,3}$ are charged. Singlet scalar charges are chosen to be $x, 2x$ respectively. Since $L_{\mu}-L_{\tau}$ gauge boson mass will be proportional to $x$ as well, in addition to $g_{\mu \tau}$ and singlet VEVs, one can have symmetry breaking scale above the scale of leptogenesis while having $g_{\mu \tau}-M_{Z_{\mu \tau}}$ in the desired range for explaining muon $(g-2)$ by choosing small $x$.

With this new particle content in table \ref{tab2}, the singlet fermion Lagrangian remain same as before (except the new gauge charges of $N_{\mu, \tau}$) while the neutrino Dirac Yukawa terms change to the following

\begin{align}
-\mathcal{L} & \supset Y_{De} \bar{L_{e}}N_{e}\tilde{H_{1}}  + Y_{D\mu}\bar{L_{\mu}}N_{\mu}\tilde{H_{2}}  +Y_{D\tau}\bar{L_{\tau}}N_{\tau}\tilde{H_{3}} + {\rm h.c.}
\end{align}

While the right handed neutrino mass matrix $M_R$ has the same structure as before, the Dirac neutrino mass matrix is changed to
\begin{equation}
\small{M_{D}}=
\begin{pmatrix}
\dfrac{Y_{De} v_{1}}{\sqrt{2}} & 0 & 0 \\
0 & \dfrac{Y_{D\mu} v_{1} \tan\beta}{\sqrt{2}} & 0 \\
0 & 0 & \dfrac{Y_{D\tau}v_{1}\tan\beta}{\sqrt{2}} 
\end{pmatrix}.
\end{equation}

Here we define $\tan \beta=v_{2}/v_{1}=v_{3}/v_{1}$, where $v_{1}$, $v_{2}$ and $v_{3}$ are the VEV of neutral components of Higgs doublets $H_{1}$, $H_{2}$ and $H_{3}$ respectively. We are assuming $v_2=v_3$ for simplicity so that $v_{1}\sqrt{1+2\tan^2\beta}=246$ GeV. While deviation from this $v_2=v_3$ assumption can lead to different phenomenology of such multi-Higgs doublet models, we do not expect leptogenesis results to change significantly and hence stick to this simple limit. Similarly, assuming the singlet VEVs to be identical to $u$, the $Z_{\mu \tau}$ mass can be derived as
\begin{equation}
M_{Z_{\mu \tau}}=\sqrt{5}x g_{\mu \tau} u.
\label{eq:Mz}
\end{equation}

The model we adopt here has two SM-singlet scalars and three Higgs doublets. Since leptogenesis occurs below the scale of $U(1)_{L_{\mu}-L_{\tau}}$ symmetry breaking, the singlet scalars are massive at the scale of leptogenesis. While they do not play any role in leptogenesis, their masses can still be constrained from experimental data. Bounds on such singlet scalars arise primarily due to its mixing with the SM like Higgs boson \cite{Robens:2015gla,Chalons:2016jeu}. The strongest bound on singlet scalar-SM like Higgs mixing angle ($\theta_m$) comes form $W$ boson mass correction \cite{Lopez-Val:2014jva} at NLO for $250 {\rm ~ GeV} \lesssim m_{s_i} \lesssim 850$ GeV as ($0.2\lesssim \sin\theta_m \lesssim 0.3$) where $m_{s_i}$ is the mass of singlet scalar $s_i$. On the other hand, for $m_{s_i}>850$ GeV, the bounds from the requirement of perturbativity and unitarity of the theory turn dominant which gives $\sin\theta_m\lesssim 0.2$. For lower values singlet masses $m_{s_i}<250$ GeV, the LHC and LEP direct search \cite{Khachatryan:2015cwa,Strassler:2006ri} and measured Higgs signal strength \cite{Strassler:2006ri} restrict the mixing angle $\sin\theta_m$ dominantly ($\lesssim 0.25$). The bounds from the measured value of EW precision parameter are mild for $m_{s_i}< 1$ TeV. Since singlet scalars do not play any role in leptogenesis, we can tune their couplings with the SM like Higgs so that the mixing angle remains as small as required. On the other hand, the physical scalars arising from Higgs doublets are also constrained from experiments. While they remain massless at the scale of leptogenesis due to unbroken electroweak symmetry, several studies like \cite{Muhlleitner:2017dkd, Haller:2018nnx, Misiak:2017bgg, Arhrib:2017uon} and references therein, have studied the phenomenology of such multi-Higgs doublet models. While the bounds from existing experiments depend upon the type of such models, the lower bounds on additional scalar masses, arising out of such additional Higgs doublets can range from around 100 GeV to a few hundred GeVs. Since we can satisfy these bounds by appropriate tuning of scalar potential parameters and without affecting the results related to leptogenesis, we do not elaborate them further in this work.

We then follow the same procedure as above by diagonalising $M_R$ and rewrite $M_D$ in the diagonal $M_R$ basis followed by solving the Boltzmann equations numerically to find the asymmetry. We keep the values of $g_{\mu \tau}$ and $M_{Z_{\mu \tau}}$ in the region favoured by muon $(g-2)$. With such small values of $g_{\mu\tau}$ and $M_{Z_{\mu \tau}}$ while keeping the singlet VEVs above the masses of the RHNs, the required value of parameter $x$ becomes less than unity reducing the interaction strength of the RHNs with $Z_{\mu \tau}$ compared to the earlier model. As a consequence, the annihilation of RHNs into a pair of $Z_{\mu \tau}$ or other annihilation mediated by $Z_{\mu \tau}$ remain sub-dominant and hence do not play any significant role in our analysis. Therefore, $N_{1,2}$ abundances are primarily determined by their decays.
\begin{figure*}
\includegraphics[scale=.5]{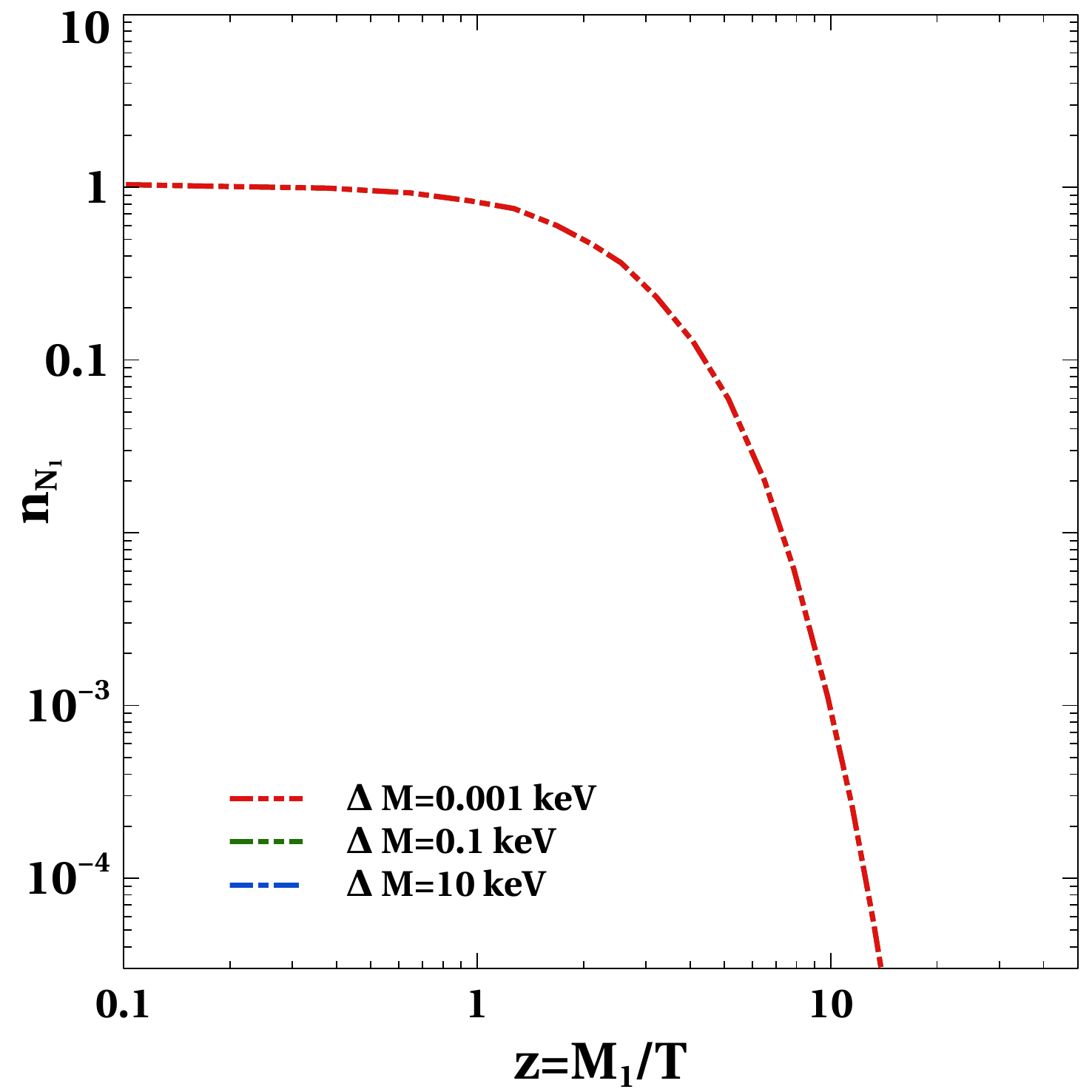}
\includegraphics[scale=.5]{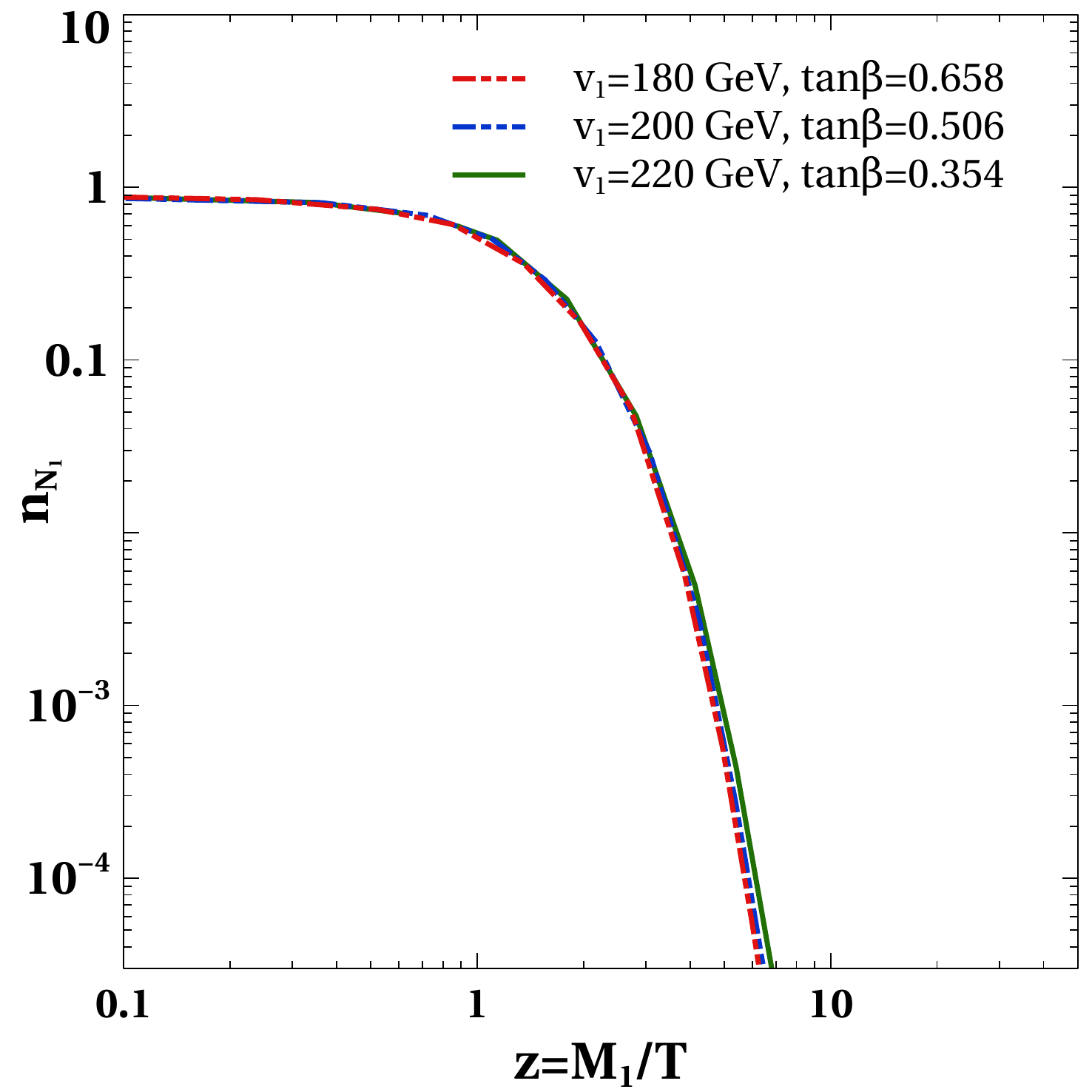}\\
\caption{Evolution plot of $N_{1}$ for different benchmark values of mass difference $\Delta M$ (left panel) and $v_{1}, \tan\beta$ (right panel). The other relevant parameters are fixed at $g_{\mu \tau}=10^{-3}$, $M_{Z_{\mu \tau}}=0.3$ GeV, $M_{1}=1.38$ TeV, $u=2$ TeV and $x=0.067$. For left panel plot $v_{1}=200$ GeV and $\tan{\beta}=0.50$ while for right panel plot $\Delta M=10$ keV.}
\label{fig:Evolution}
\end{figure*}

\begin{figure*}
\includegraphics[scale=.5]{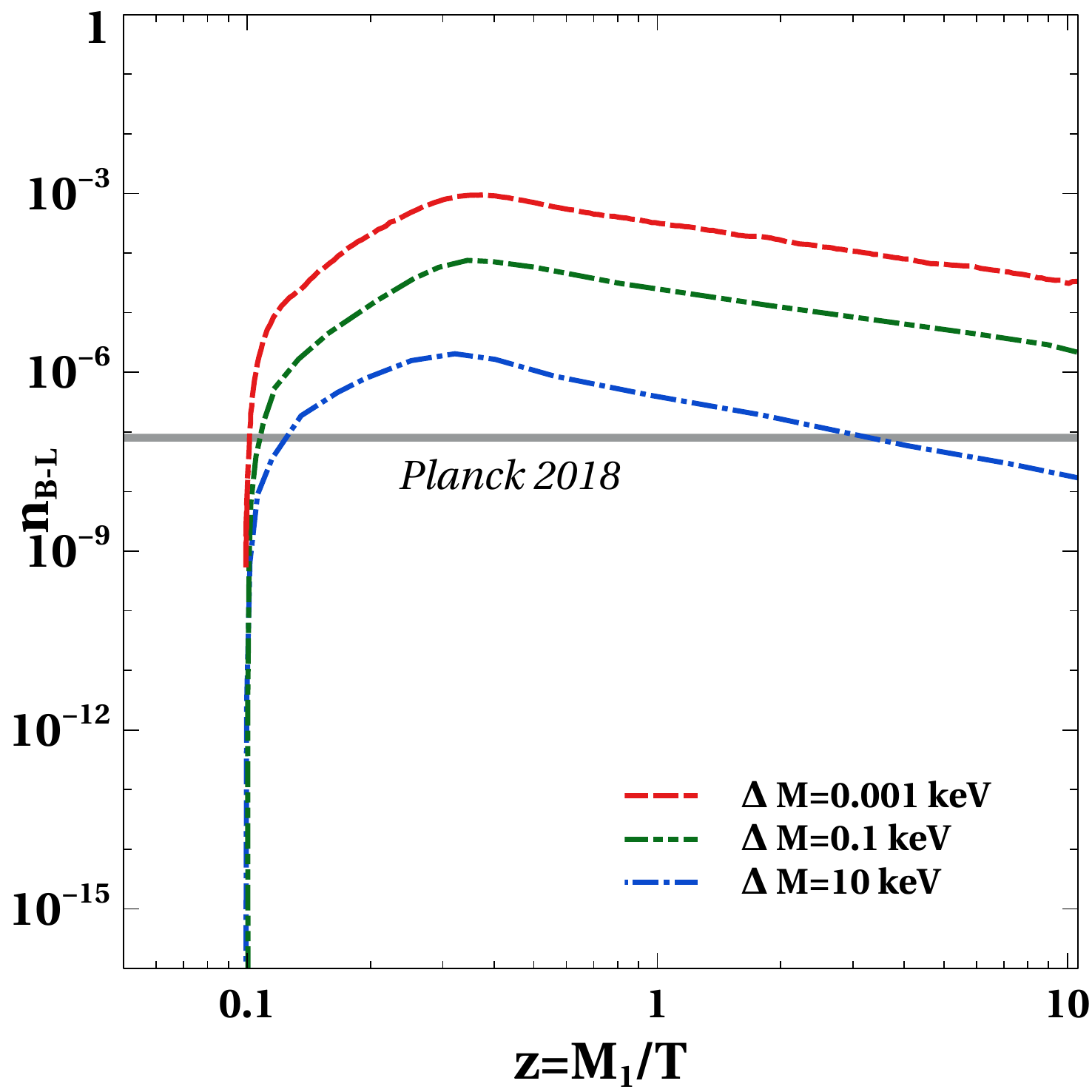}
\includegraphics[scale=.5]{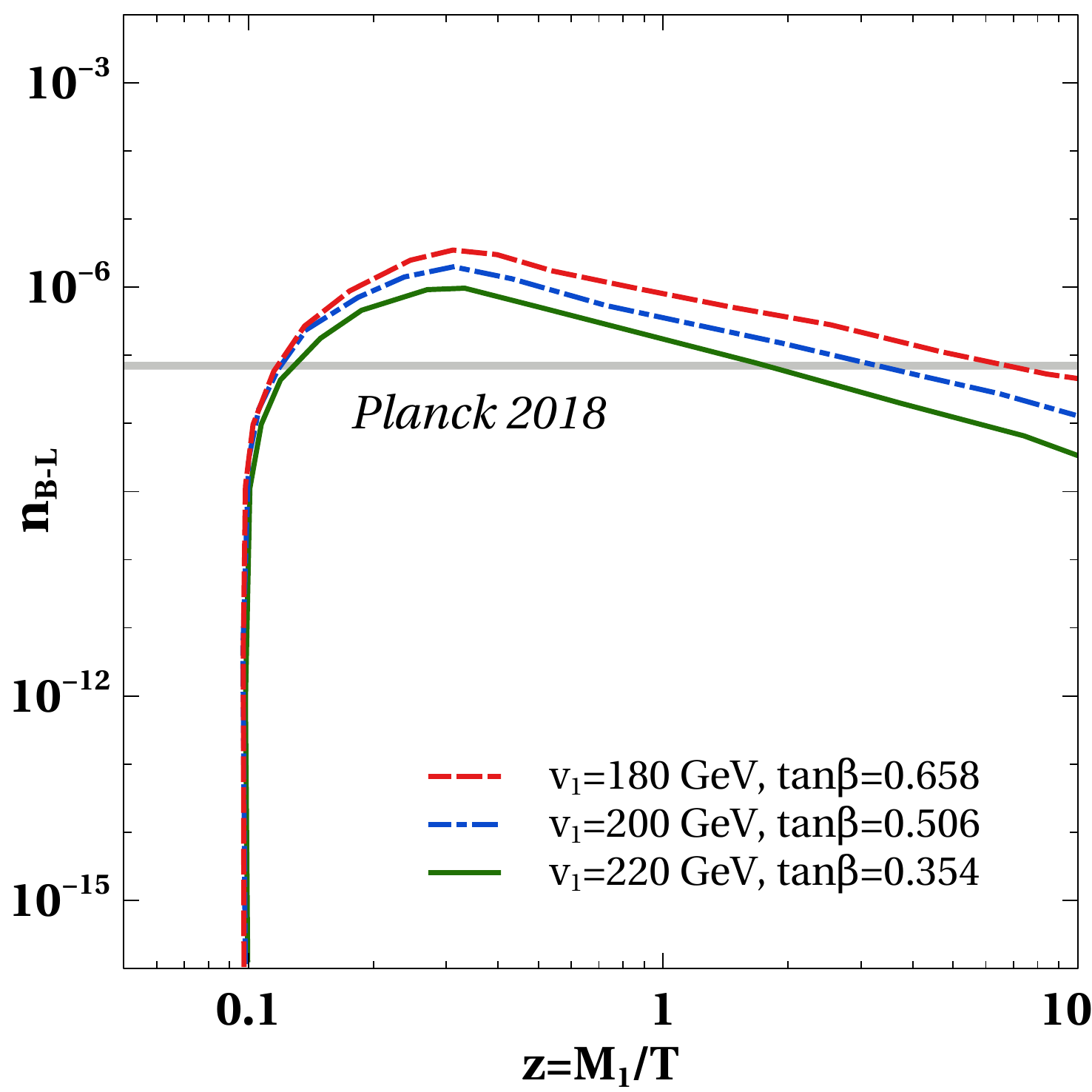} \\
\caption{Evolution plot of $B-L$ asymmetry for different possible benchmark combinations of mass difference $\Delta M$ (left panel) and $v_{1}, \tan\beta$ (right panel). The other relevant parameters are fixed at $g_{\mu \tau}=10^{-3}$, $M_{Z_{\mu \tau}}=0.3$ GeV, $M_{1}=1.38$ TeV, $u=2$ TeV and $x=0.067$. For left panel plot $v_{1}=200$ GeV and $\tan{\beta}=0.50$ while for right panel plot $\Delta M=10$ keV.}
\label{fig:Evolution2}
\end{figure*}
Similarly, in the muon $(g-2)$ favoured regime of $g_{\mu\tau}-M_{Z_{\mu \tau}}$, the washouts involving $Z_{\mu \tau}$ are very feeble and do not produce any significant effect on asymmetry. The dominant washout processes in this regime are the inverse decays and the scattering $q l \longrightarrow q N_{1,2}$. In figure \ref{fig:Evolution} we show the evolution of $N_1$ abundance for different benchmark parameters. Since $N_1$ abundance is primarily governed by decays in this regime, we do not see much differences due to change in model parameters like mass splitting $\Delta M$ or $\tan \beta$. Similarly, in figure \ref{fig:Evolution2}, we show the evolution of $n_{B-L}$ for different combination of mass splitting $\Delta M$ (left panel) and $tan\beta$ (right panel). Clearly, the asymmetry is maximum for smallest mass splitting. It is because of the resonance enhancement of the asymmetry. The decay widths $\Gamma_{1,2}$ are of the order of $10^{-10}-10^{-9}$ GeV, within the parameter space we are interested in and as the mass splitting reaches near that value the CP asymmetry increases sharply to 
$\mathcal{O}(1)$ due to the resonance enhancement condition $\Delta M \sim \Gamma/2$ mentioned earlier. On the right panel of figure \ref{fig:Evolution2} it is seen that asymmetry increases for larger values of $v_{1}\tan\beta$. Once we fix the entries of the Dirac mass matrix from neutrino oscillation data, the Yukawa couplings ($Y_{D\mu}$ and $Y_{D\tau}$) relevant for leptogensis are very tightly constrained to $v_{1}tan\beta$. Since $v_{1}\tan\beta = v_2=v_3$, one requires smaller values of Yukawa couplings ($Y_{D\mu}$ and $Y_{D\tau}$) for larger values of $v_{1}tan\beta$ to satisfy light neutrino data, for a fixed scale of leptogenesis. As a consequence of that the decay widths of $N_{1,2}$ decrease and also the corresponding inverse decay rates. The small difference in the decay rates of $N_{1}$ is marginally visible in the evolution plots of $n_{N_{1,2}}$ of figure \ref{fig:Evolution} (right panel) as well. Since lepton asymmetry is generated primarily due to resonant enhancement, decrease in Yukawa coupling (or increase in $v_{1}tan\beta$) decreases the inverse decay rates which increases the $B-L$ asymmetry as can be seen from the right panel plot of figure \ref{fig:Evolution2}. 

\begin{figure}
\includegraphics[scale=.5]{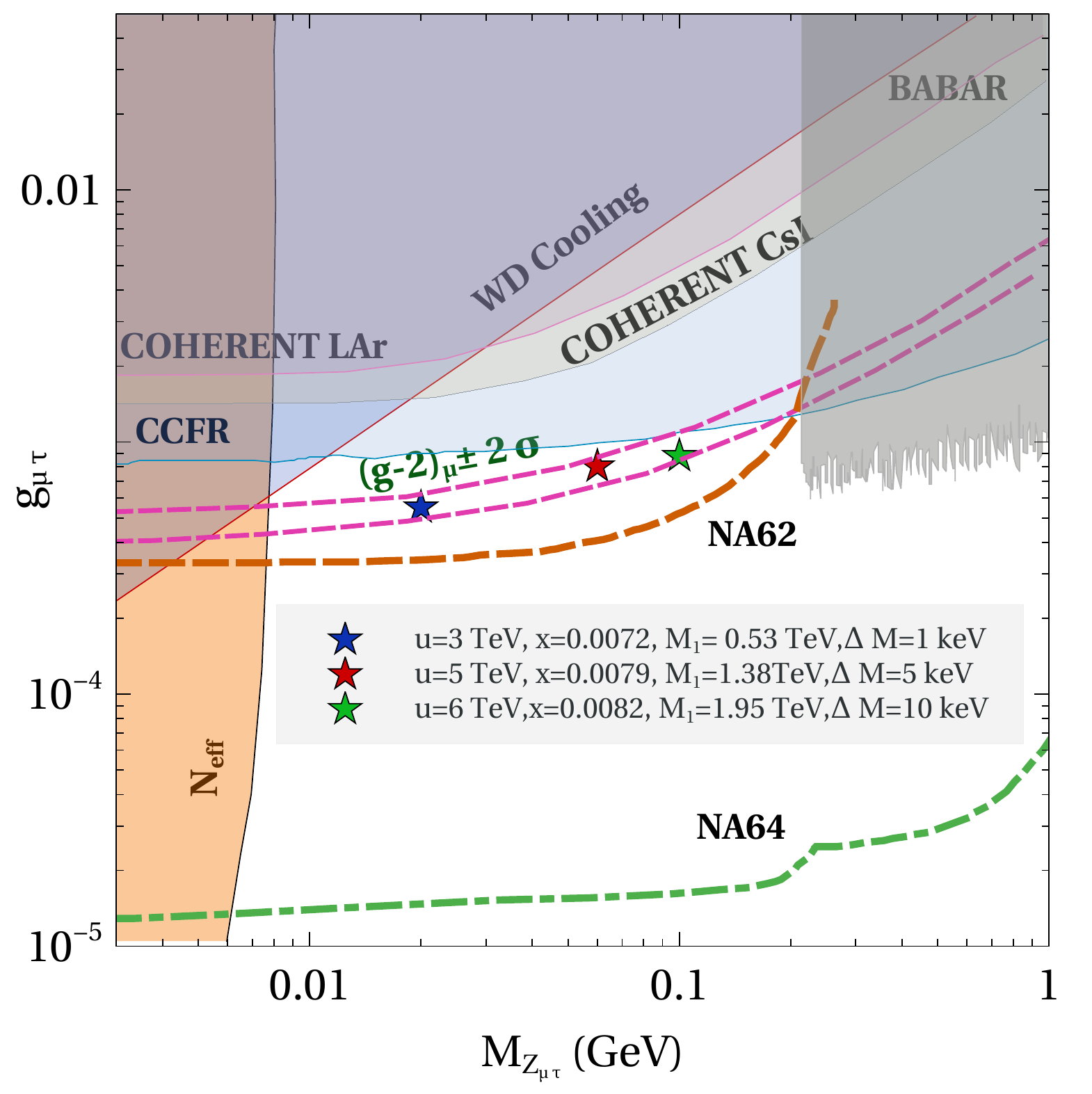}
\caption{Summary plot showing the allowed parameter space in $g_{\mu \tau}-M_{Z_{\mu \tau}}$ plane for sub-GeV $M_{Z_{\mu \tau}}$. The region between pink dashed lines represents muon $(g-2)$ favoured region, a small portion of which remain allowed from different exclusion limits shown as shaded regions. The points marked as stars denote benchmark points satisfying the criteria of successful TeV scale resonant leptogenesis (see text for details).}
\label{Fig2}
\end{figure}

Due to the feeble gauge portal interactions in the low mass regime of $Z_{\mu \tau}$, we do not find any strong correlation in $g_{\mu \tau}-M_{\mu \tau}$ plane from leptogenesis unlike in the previous model. Nevertheless, it is observed that low scale leptogenesis is possible in the region of $g_{\mu \tau}-M_{Z_{\mu \tau}}$ plane favoured by muon ($g-2$). We show a summary plot in figure \ref{Fig2} showing the allowed parameter space in $g_{\mu \tau}-M_{Z_{\mu \tau}}$ plane and show three possible benchmark points for TeV scale leptogenesis which also satisfy the light neutrino data. 

In equation \eqref{eq:BenchmarkYuk} we show the numerical value of the Yukawa matrix $h$ for first benchmark point of figure \ref{Fig2}.

\begin{widetext}
\begin{equation}\label{eq:BenchmarkYuk}
h \simeq 10^{-7} \times \begin{pmatrix}
0.897+3.138i &  9.737-8.179i  & -1.164-0.8033i \\
3.204+4.101i & -6.792-4.447i & 3.460-5.711i \\
-0.1577+3.778i & -0.885-5.452i & 2.639+9.414i
\end{pmatrix}
\end{equation}
\end{widetext}

Since the viability of resonant leptogenesis crucially depends upon quasi-degenerate right handed neutrinos, we show the structure of the RHN mass matrix too corresponding to the first benchmark point of figure \ref{Fig2} as
\begin{equation}\label{eq:BenchmarkMR}
M_R \simeq \begin{pmatrix}
0.602 &  0.106 (1+i)  & 0.106(1-i) \\
0.106 (1+i) & 0.424(1-i) & 0.125 \\
0.106(1-i) & 0.125 & 0.636(1+i)
\end{pmatrix}
\end{equation}
in the units of TeV. Accordingly, the bare mass terms $M_{ee}, M_{\mu \tau}$ as well as the singlet scalar Yukawa couplings with the RHNs can be tuned. The above mass matrix gives rise to two quasi-degenerate RHNs with masses $M_1 = M_2-\Delta M =0.53$ TeV with mass splitting $\Delta M= 1$ keV and $M_3=1.6$ TeV.

In summary plot of figure \ref{Fig2}, the parameter space corresponding to Fermilab's muon $(g-2)$ data is shown between the dashed pink coloured lines. Several experimental constraints are also shown in the summary plot of figure \ref{Fig2}. The cyan coloured exclusion band labelled as CCFR correspond to the upper bound on neutrino trident process measured by the CCFR collaboration \cite{Altmannshofer:2014pba}. In the high mass regime for $Z_{\mu \tau}$ the exclusion region labelled as BABAR corresponds to the limits imposed on four muon final states by the BABAR collaboration \cite{TheBABAR:2016rlg}. The astrophysical bounds from cooling of white dwarf (WD) \cite{Bauer:2018onh, Kamada:2018zxi} excludes the upper left triangular region. Very light $Z_{\mu \tau}$ is ruled out from cosmological constraints on effective relativistic degrees of freedom \cite{Aghanim:2018eyx, Kamada:2018zxi, Ibe:2019gpv, Escudero:2019gzq}. This arises due to the late decay of such light gauge bosons into SM leptons, after standard neutrino decoupling temperatures thereby enhancing $N_{\rm eff}$. More stringent constraints apply if DM is in the sub-GeV regime and its thermal relic is dictated by annihilation mediated by $Z_{\mu \tau}$. However, in our case, such additional constraints do not arise as DM is much heavier and uncharged under $L_{\mu}-L_{\tau}$ gauge symmetry. The observation of coherent elastic neutrino-nucleus cross section in liquid argon (LAr) and cesium-iodide (CsI) performed by the COHERENT Collaboration \cite{Akimov:2017ade, Akimov:2020pdx} also leads to constraint on $g_{\mu \tau}-M_{Z_{\mu \tau}}$ parameter space. Adopting the analysis of \cite{Cadeddu:2020nbr, Banerjee:2021laz}, we impose these bounds, labelled as COHERENT LAr and COHERENT CsI respectively in figure \ref{Fig2}. Clearly, even after incorporating all existing experimental bounds, there still exists a small parameter space between a few MeV to around 100 MeV consistent with all bounds and the requirement of explaining muon $(g-2)$. The three points denoted by stars correspond to benchmark values of parameters for which the correct baryon asymmetry can be generated via resonant leptogenesis. Therefore, we can have successful TeV scale resonant leptogenesis while also satisfying the Fermilab's data on muon (g-2). This allowed region from muon (g-2) and also TeV scale leptogenesis remains within the reach of future experiments like NA62 at CERN \cite{Krnjaic:2019rsv} (Orange dashed line in figure~\ref{Fig2}), the NA64 experiment at CERN (dot-dashed line of green colour in  figure~\ref{Fig2}) \cite{Gninenko:2014pea, Gninenko:2018tlp}. Possible future confirmation of this muon $g-2$ favoured parameter space will also indicate the possible scale of leptogenesis in this model. \\

\noindent
{\bf Conclusion}: 
We have studied the possibility of TeV scale resonant leptogenesis in gauged $L_{\mu}-L_{\tau}$ model with type I seesaw origin of light neutrino masses. While the minimal version of such a model has only one scalar singlet, we found it too constrained from light neutrino data specially if we demand two of the right handed neutrinos to be quasi-degenerate from the requirement of resonant leptogenesis. We then considered the two singlet scalars version of this model and showed the possibility of TeV scale leptogenesis while satisfying light neutrino data. We also found interesting correlations between $L_{\mu}-L_{\tau}$ gauge sector parameters from the criteria of successful leptogenesis. Since the requirement of successful leptogenesis at TeV scale requires broken $L_{\mu}-L_{\tau}$ gauge symmetry, this minimal setup requires heavy $Z_{\mu \tau}$ gauge boson above GeV scale. On the other hand, the muon $(g-2)$ favoured parameter space for such heavy $Z_{\mu \tau}$ is already ruled out by neutrino trident data and (partially) by the LHC constraints on $Z \rightarrow 4 \mu$. We then consider an extended version of this model with the possibility of sub-GeV $Z_{\mu \tau}$ even with TeV scale $L_{\mu}-L_{\tau}$ gauge symmetry breaking due to singlet scalars having fractional gauge charge less than unity. This requires two right handed neutrinos to have fractional charges too requiring the presence of additional Higgs doublets to allow the required Dirac neutrino Yukawa couplings. We show that, successful TeV scale resonant leptogenesis is possible in this model while satisfying muon $(g-2)$ and light neutrino data and also evading other experimental constraints on such sub-GeV leptophillic gauge sector. 
The presence of additional Higgs doublets can also give rise to interesting phenomenology, similar to two Higgs doublet models discussed extensively in the literature \cite{Branco:2011iw}.

\noindent
\acknowledgements
DB acknowledges the support from Early Career Research Award from DST-SERB, Government of India (reference number: ECR/2017/001873).

\begin{thebibliography}{0}
\expandafter\ifx\csname natexlab\endcsname\relax\def\natexlab#1{#1}\fi
\expandafter\ifx\csname bibnamefont\endcsname\relax
  \def\bibnamefont#1{#1}\fi
\expandafter\ifx\csname bibfnamefont\endcsname\relax
  \def\bibfnamefont#1{#1}\fi
\expandafter\ifx\csname citenamefont\endcsname\relax
  \def\citenamefont#1{#1}\fi
\expandafter\ifx\csname url\endcsname\relax
  \def\url#1{\texttt{#1}}\fi
\expandafter\ifx\csname urlprefix\endcsname\relax\def\urlprefix{URL }\fi
\providecommand{\bibinfo}[2]{#2}
\providecommand{\eprint}[2][]{\url{#2}}

\end{thebibliography}


\begin{thebibliography}{75}
\expandafter\ifx\csname natexlab\endcsname\relax\def\natexlab#1{#1}\fi
\expandafter\ifx\csname bibnamefont\endcsname\relax
  \def\bibnamefont#1{#1}\fi
\expandafter\ifx\csname bibfnamefont\endcsname\relax
  \def\bibfnamefont#1{#1}\fi
\expandafter\ifx\csname citenamefont\endcsname\relax
  \def\citenamefont#1{#1}\fi
\expandafter\ifx\csname url\endcsname\relax
  \def\url#1{\texttt{#1}}\fi
\expandafter\ifx\csname urlprefix\endcsname\relax\def\urlprefix{URL }\fi
\providecommand{\bibinfo}[2]{#2}
\providecommand{\eprint}[2][]{\url{#2}}

\bibitem[{\citenamefont{Abi et~al.}(2021)}]{Abi:2021gix}
\bibinfo{author}{\bibfnamefont{B.}~\bibnamefont{Abi}} \bibnamefont{et~al.}
  (\bibinfo{collaboration}{Muon g-2}), \bibinfo{journal}{Phys. Rev. Lett.}
  \textbf{\bibinfo{volume}{126}}, \bibinfo{pages}{141801}
  (\bibinfo{year}{2021}), \eprint{2104.03281}.

\bibitem[{\citenamefont{Borsanyi et~al.}(2020)}]{Borsanyi:2020mff}
\bibinfo{author}{\bibfnamefont{S.}~\bibnamefont{Borsanyi}} \bibnamefont{et~al.}
  (\bibinfo{year}{2020}), \eprint{2002.12347}.

\bibitem[{\citenamefont{Crivellin et~al.}(2020)\citenamefont{Crivellin,
  Hoferichter, Manzari, and Montull}}]{Crivellin:2020zul}
\bibinfo{author}{\bibfnamefont{A.}~\bibnamefont{Crivellin}},
  \bibinfo{author}{\bibfnamefont{M.}~\bibnamefont{Hoferichter}},
  \bibinfo{author}{\bibfnamefont{C.~A.} \bibnamefont{Manzari}},
  \bibnamefont{and} \bibinfo{author}{\bibfnamefont{M.}~\bibnamefont{Montull}},
  \bibinfo{journal}{Phys. Rev. Lett.} \textbf{\bibinfo{volume}{125}},
  \bibinfo{pages}{091801} (\bibinfo{year}{2020}), \eprint{2003.04886}.

\bibitem[{\citenamefont{Aoyama et~al.}(2020)}]{Aoyama:2020ynm}
\bibinfo{author}{\bibfnamefont{T.}~\bibnamefont{Aoyama}} \bibnamefont{et~al.}
  (\bibinfo{year}{2020}), \eprint{2006.04822}.

\bibitem[{\citenamefont{Zyla et~al.}(2020)}]{Zyla:2020zbs}
\bibinfo{author}{\bibfnamefont{P.~A.} \bibnamefont{Zyla}} \bibnamefont{et~al.}
  (\bibinfo{collaboration}{Particle Data Group}), \bibinfo{journal}{PTEP}
  \textbf{\bibinfo{volume}{2020}}, \bibinfo{pages}{083C01}
  (\bibinfo{year}{2020}).

\bibitem[{\citenamefont{Athron et~al.}(2021)\citenamefont{Athron, Bal\'azs,
  Jacob, Kotlarski, St\"ockinger, and St\"ockinger-Kim}}]{Athron:2021iuf}
\bibinfo{author}{\bibfnamefont{P.}~\bibnamefont{Athron}},
  \bibinfo{author}{\bibfnamefont{C.}~\bibnamefont{Bal\'azs}},
  \bibinfo{author}{\bibfnamefont{D.~H.} \bibnamefont{Jacob}},
  \bibinfo{author}{\bibfnamefont{W.}~\bibnamefont{Kotlarski}},
  \bibinfo{author}{\bibfnamefont{D.}~\bibnamefont{St\"ockinger}},
  \bibnamefont{and}
  \bibinfo{author}{\bibfnamefont{H.}~\bibnamefont{St\"ockinger-Kim}}
  (\bibinfo{year}{2021}), \eprint{2104.03691}.

\bibitem[{\citenamefont{Borah et~al.}(2020)\citenamefont{Borah, Mahapatra,
  Nanda, and Sahu}}]{Borah:2020jzi}
\bibinfo{author}{\bibfnamefont{D.}~\bibnamefont{Borah}},
  \bibinfo{author}{\bibfnamefont{S.}~\bibnamefont{Mahapatra}},
  \bibinfo{author}{\bibfnamefont{D.}~\bibnamefont{Nanda}}, \bibnamefont{and}
  \bibinfo{author}{\bibfnamefont{N.}~\bibnamefont{Sahu}}
  (\bibinfo{year}{2020}), \eprint{2007.10754}.

\bibitem[{\citenamefont{Zu et~al.}(2021)\citenamefont{Zu, Pan, Feng, Yuan, and
  Fan}}]{Zu:2021odn}
\bibinfo{author}{\bibfnamefont{L.}~\bibnamefont{Zu}},
  \bibinfo{author}{\bibfnamefont{X.}~\bibnamefont{Pan}},
  \bibinfo{author}{\bibfnamefont{L.}~\bibnamefont{Feng}},
  \bibinfo{author}{\bibfnamefont{Q.}~\bibnamefont{Yuan}}, \bibnamefont{and}
  \bibinfo{author}{\bibfnamefont{Y.-Z.} \bibnamefont{Fan}}
  (\bibinfo{year}{2021}), \eprint{2104.03340}.

\bibitem[{\citenamefont{Amaral et~al.}(2021)\citenamefont{Amaral, Cerde\~no,
  Cheek, and Foldenauer}}]{Amaral:2021rzw}
\bibinfo{author}{\bibfnamefont{D.~W.~P.} \bibnamefont{Amaral}},
  \bibinfo{author}{\bibfnamefont{D.~G.} \bibnamefont{Cerde\~no}},
  \bibinfo{author}{\bibfnamefont{A.}~\bibnamefont{Cheek}}, \bibnamefont{and}
  \bibinfo{author}{\bibfnamefont{P.}~\bibnamefont{Foldenauer}}
  (\bibinfo{year}{2021}), \eprint{2104.03297}.

\bibitem[{\citenamefont{Zhou}(2021)}]{Zhou:2021vnf}
\bibinfo{author}{\bibfnamefont{S.}~\bibnamefont{Zhou}} (\bibinfo{year}{2021}),
  \eprint{2104.06858}.

\bibitem[{\citenamefont{Borah et~al.}(2021)\citenamefont{Borah, Dutta,
  Mahapatra, and Sahu}}]{Borah:2021jzu}
\bibinfo{author}{\bibfnamefont{D.}~\bibnamefont{Borah}},
  \bibinfo{author}{\bibfnamefont{M.}~\bibnamefont{Dutta}},
  \bibinfo{author}{\bibfnamefont{S.}~\bibnamefont{Mahapatra}},
  \bibnamefont{and} \bibinfo{author}{\bibfnamefont{N.}~\bibnamefont{Sahu}}
  (\bibinfo{year}{2021}), \eprint{2104.05656}.

\bibitem[{\citenamefont{Aaij et~al.}(2021)}]{Aaij:2021vac}
\bibinfo{author}{\bibfnamefont{R.}~\bibnamefont{Aaij}} \bibnamefont{et~al.}
  (\bibinfo{collaboration}{LHCb}) (\bibinfo{year}{2021}), \eprint{2103.11769}.

\bibitem[{\citenamefont{He et~al.}(1991{\natexlab{a}})\citenamefont{He, Joshi,
  Lew, and Volkas}}]{He:1990pn}
\bibinfo{author}{\bibfnamefont{X.}~\bibnamefont{He}},
  \bibinfo{author}{\bibfnamefont{G.~C.} \bibnamefont{Joshi}},
  \bibinfo{author}{\bibfnamefont{H.}~\bibnamefont{Lew}}, \bibnamefont{and}
  \bibinfo{author}{\bibfnamefont{R.}~\bibnamefont{Volkas}},
  \bibinfo{journal}{Phys. Rev. D} \textbf{\bibinfo{volume}{43}},
  \bibinfo{pages}{22} (\bibinfo{year}{1991}{\natexlab{a}}).

\bibitem[{\citenamefont{He et~al.}(1991{\natexlab{b}})\citenamefont{He, Joshi,
  Lew, and Volkas}}]{He:1991qd}
\bibinfo{author}{\bibfnamefont{X.-G.} \bibnamefont{He}},
  \bibinfo{author}{\bibfnamefont{G.~C.} \bibnamefont{Joshi}},
  \bibinfo{author}{\bibfnamefont{H.}~\bibnamefont{Lew}}, \bibnamefont{and}
  \bibinfo{author}{\bibfnamefont{R.~R.} \bibnamefont{Volkas}},
  \bibinfo{journal}{Phys. Rev. D} \textbf{\bibinfo{volume}{44}},
  \bibinfo{pages}{2118} (\bibinfo{year}{1991}{\natexlab{b}}).

\bibitem[{\citenamefont{Minkowski}(1977)}]{Minkowski:1977sc}
\bibinfo{author}{\bibfnamefont{P.}~\bibnamefont{Minkowski}},
  \bibinfo{journal}{Phys. Lett. B} \textbf{\bibinfo{volume}{67}},
  \bibinfo{pages}{421} (\bibinfo{year}{1977}).

\bibitem[{\citenamefont{Mohapatra and Senjanovic}(1980)}]{Mohapatra:1979ia}
\bibinfo{author}{\bibfnamefont{R.~N.} \bibnamefont{Mohapatra}}
  \bibnamefont{and}
  \bibinfo{author}{\bibfnamefont{G.}~\bibnamefont{Senjanovic}},
  \bibinfo{journal}{Phys. Rev. Lett.} \textbf{\bibinfo{volume}{44}},
  \bibinfo{pages}{912} (\bibinfo{year}{1980}).

\bibitem[{\citenamefont{Yanagida}(1979)}]{Yanagida:1979as}
\bibinfo{author}{\bibfnamefont{T.}~\bibnamefont{Yanagida}},
  \bibinfo{journal}{Conf. Proc.} \textbf{\bibinfo{volume}{C7902131}},
  \bibinfo{pages}{95} (\bibinfo{year}{1979}).

\bibitem[{\citenamefont{Gell-Mann et~al.}(1979)\citenamefont{Gell-Mann, Ramond,
  and Slansky}}]{GellMann:1980vs}
\bibinfo{author}{\bibfnamefont{M.}~\bibnamefont{Gell-Mann}},
  \bibinfo{author}{\bibfnamefont{P.}~\bibnamefont{Ramond}}, \bibnamefont{and}
  \bibinfo{author}{\bibfnamefont{R.}~\bibnamefont{Slansky}},
  \bibinfo{journal}{Conf. Proc. C} \textbf{\bibinfo{volume}{790927}},
  \bibinfo{pages}{315} (\bibinfo{year}{1979}), \eprint{1306.4669}.

\bibitem[{\citenamefont{Glashow}(1980)}]{Glashow:1979nm}
\bibinfo{author}{\bibfnamefont{S.}~\bibnamefont{Glashow}},
  \bibinfo{journal}{NATO Sci. Ser. B} \textbf{\bibinfo{volume}{61}},
  \bibinfo{pages}{687} (\bibinfo{year}{1980}).

\bibitem[{\citenamefont{Schechter and Valle}(1980)}]{Schechter:1980gr}
\bibinfo{author}{\bibfnamefont{J.}~\bibnamefont{Schechter}} \bibnamefont{and}
  \bibinfo{author}{\bibfnamefont{J.}~\bibnamefont{Valle}},
  \bibinfo{journal}{Phys. Rev. D} \textbf{\bibinfo{volume}{22}},
  \bibinfo{pages}{2227} (\bibinfo{year}{1980}).

\bibitem[{\citenamefont{Fukugita and Yanagida}(1986)}]{Fukugita:1986hr}
\bibinfo{author}{\bibfnamefont{M.}~\bibnamefont{Fukugita}} \bibnamefont{and}
  \bibinfo{author}{\bibfnamefont{T.}~\bibnamefont{Yanagida}},
  \bibinfo{journal}{Phys. Lett.} \textbf{\bibinfo{volume}{B174}},
  \bibinfo{pages}{45} (\bibinfo{year}{1986}).

\bibitem[{\citenamefont{Davidson et~al.}(2008)\citenamefont{Davidson, Nardi,
  and Nir}}]{Davidson:2008bu}
\bibinfo{author}{\bibfnamefont{S.}~\bibnamefont{Davidson}},
  \bibinfo{author}{\bibfnamefont{E.}~\bibnamefont{Nardi}}, \bibnamefont{and}
  \bibinfo{author}{\bibfnamefont{Y.}~\bibnamefont{Nir}},
  \bibinfo{journal}{Phys. Rept.} \textbf{\bibinfo{volume}{466}},
  \bibinfo{pages}{105} (\bibinfo{year}{2008}), \eprint{0802.2962}.

\bibitem[{\citenamefont{Davidson and Ibarra}(2002)}]{Davidson:2002qv}
\bibinfo{author}{\bibfnamefont{S.}~\bibnamefont{Davidson}} \bibnamefont{and}
  \bibinfo{author}{\bibfnamefont{A.}~\bibnamefont{Ibarra}},
  \bibinfo{journal}{Phys. Lett.} \textbf{\bibinfo{volume}{B535}},
  \bibinfo{pages}{25} (\bibinfo{year}{2002}), \eprint{hep-ph/0202239}.

\bibitem[{\citenamefont{Adhikary}(2006)}]{Adhikary:2006rf}
\bibinfo{author}{\bibfnamefont{B.}~\bibnamefont{Adhikary}},
  \bibinfo{journal}{Phys. Rev. D} \textbf{\bibinfo{volume}{74}},
  \bibinfo{pages}{033002} (\bibinfo{year}{2006}), \eprint{hep-ph/0604009}.

\bibitem[{\citenamefont{Chun and Turzynski}(2007)}]{Chun:2007vh}
\bibinfo{author}{\bibfnamefont{E.~J.} \bibnamefont{Chun}} \bibnamefont{and}
  \bibinfo{author}{\bibfnamefont{K.}~\bibnamefont{Turzynski}},
  \bibinfo{journal}{Phys. Rev. D} \textbf{\bibinfo{volume}{76}},
  \bibinfo{pages}{053008} (\bibinfo{year}{2007}), \eprint{hep-ph/0703070}.

\bibitem[{\citenamefont{Ota and Rodejohann}(2006)}]{Ota:2006xr}
\bibinfo{author}{\bibfnamefont{T.}~\bibnamefont{Ota}} \bibnamefont{and}
  \bibinfo{author}{\bibfnamefont{W.}~\bibnamefont{Rodejohann}},
  \bibinfo{journal}{Phys. Lett. B} \textbf{\bibinfo{volume}{639}},
  \bibinfo{pages}{322} (\bibinfo{year}{2006}), \eprint{hep-ph/0605231}.

\bibitem[{\citenamefont{Asai et~al.}(2017)\citenamefont{Asai, Hamaguchi, and
  Nagata}}]{Asai:2017ryy}
\bibinfo{author}{\bibfnamefont{K.}~\bibnamefont{Asai}},
  \bibinfo{author}{\bibfnamefont{K.}~\bibnamefont{Hamaguchi}},
  \bibnamefont{and} \bibinfo{author}{\bibfnamefont{N.}~\bibnamefont{Nagata}},
  \bibinfo{journal}{Eur. Phys. J. C} \textbf{\bibinfo{volume}{77}},
  \bibinfo{pages}{763} (\bibinfo{year}{2017}), \eprint{1705.00419}.

\bibitem[{\citenamefont{Asai et~al.}(2020)\citenamefont{Asai, Hamaguchi,
  Nagata, and Tseng}}]{Asai:2020qax}
\bibinfo{author}{\bibfnamefont{K.}~\bibnamefont{Asai}},
  \bibinfo{author}{\bibfnamefont{K.}~\bibnamefont{Hamaguchi}},
  \bibinfo{author}{\bibfnamefont{N.}~\bibnamefont{Nagata}}, \bibnamefont{and}
  \bibinfo{author}{\bibfnamefont{S.-Y.} \bibnamefont{Tseng}},
  \bibinfo{journal}{JCAP} \textbf{\bibinfo{volume}{11}}, \bibinfo{pages}{013}
  (\bibinfo{year}{2020}), \eprint{2005.01039}.

\bibitem[{\citenamefont{Pilaftsis}(1999)}]{Pilaftsis:1998pd}
\bibinfo{author}{\bibfnamefont{A.}~\bibnamefont{Pilaftsis}},
  \bibinfo{journal}{Int. J. Mod. Phys.} \textbf{\bibinfo{volume}{A14}},
  \bibinfo{pages}{1811} (\bibinfo{year}{1999}), \eprint{hep-ph/9812256}.

\bibitem[{\citenamefont{Pilaftsis and Underwood}(2004)}]{Pilaftsis:2003gt}
\bibinfo{author}{\bibfnamefont{A.}~\bibnamefont{Pilaftsis}} \bibnamefont{and}
  \bibinfo{author}{\bibfnamefont{T.~E.~J.} \bibnamefont{Underwood}},
  \bibinfo{journal}{Nucl. Phys.} \textbf{\bibinfo{volume}{B692}},
  \bibinfo{pages}{303} (\bibinfo{year}{2004}), \eprint{hep-ph/0309342}.

\bibitem[{\citenamefont{Moffat et~al.}(2018)\citenamefont{Moffat, Pascoli,
  Petcov, Schulz, and Turner}}]{Moffat:2018wke}
\bibinfo{author}{\bibfnamefont{K.}~\bibnamefont{Moffat}},
  \bibinfo{author}{\bibfnamefont{S.}~\bibnamefont{Pascoli}},
  \bibinfo{author}{\bibfnamefont{S.~T.} \bibnamefont{Petcov}},
  \bibinfo{author}{\bibfnamefont{H.}~\bibnamefont{Schulz}}, \bibnamefont{and}
  \bibinfo{author}{\bibfnamefont{J.}~\bibnamefont{Turner}},
  \bibinfo{journal}{Phys. Rev.} \textbf{\bibinfo{volume}{D98}},
  \bibinfo{pages}{015036} (\bibinfo{year}{2018}), \eprint{1804.05066}.

\bibitem[{\citenamefont{Dev et~al.}(2018{\natexlab{a}})\citenamefont{Dev,
  Garny, Klaric, Millington, and Teresi}}]{Dev:2017wwc}
\bibinfo{author}{\bibfnamefont{P.~S.~B.} \bibnamefont{Dev}},
  \bibinfo{author}{\bibfnamefont{M.}~\bibnamefont{Garny}},
  \bibinfo{author}{\bibfnamefont{J.}~\bibnamefont{Klaric}},
  \bibinfo{author}{\bibfnamefont{P.}~\bibnamefont{Millington}},
  \bibnamefont{and} \bibinfo{author}{\bibfnamefont{D.}~\bibnamefont{Teresi}},
  \bibinfo{journal}{Int. J. Mod. Phys.} \textbf{\bibinfo{volume}{A33}},
  \bibinfo{pages}{1842003} (\bibinfo{year}{2018}{\natexlab{a}}),
  \eprint{1711.02863}.

\bibitem[{\citenamefont{Biswas and Shaw}(2019)}]{Biswas:2019twf}
\bibinfo{author}{\bibfnamefont{A.}~\bibnamefont{Biswas}} \bibnamefont{and}
  \bibinfo{author}{\bibfnamefont{A.}~\bibnamefont{Shaw}},
  \bibinfo{journal}{JHEP} \textbf{\bibinfo{volume}{05}}, \bibinfo{pages}{165}
  (\bibinfo{year}{2019}), \eprint{1903.08745}.

\bibitem[{\citenamefont{Crivellin et~al.}(2015)\citenamefont{Crivellin,
  D'Ambrosio, and Heeck}}]{Crivellin:2015mga}
\bibinfo{author}{\bibfnamefont{A.}~\bibnamefont{Crivellin}},
  \bibinfo{author}{\bibfnamefont{G.}~\bibnamefont{D'Ambrosio}},
  \bibnamefont{and} \bibinfo{author}{\bibfnamefont{J.}~\bibnamefont{Heeck}},
  \bibinfo{journal}{Phys. Rev. Lett.} \textbf{\bibinfo{volume}{114}},
  \bibinfo{pages}{151801} (\bibinfo{year}{2015}), \eprint{1501.00993}.

\bibitem[{\citenamefont{Altmannshofer et~al.}(2016)\citenamefont{Altmannshofer,
  Carena, and Crivellin}}]{Altmannshofer:2016oaq}
\bibinfo{author}{\bibfnamefont{W.}~\bibnamefont{Altmannshofer}},
  \bibinfo{author}{\bibfnamefont{M.}~\bibnamefont{Carena}}, \bibnamefont{and}
  \bibinfo{author}{\bibfnamefont{A.}~\bibnamefont{Crivellin}},
  \bibinfo{journal}{Phys. Rev. D} \textbf{\bibinfo{volume}{94}},
  \bibinfo{pages}{095026} (\bibinfo{year}{2016}), \eprint{1604.08221}.

\bibitem[{\citenamefont{Esteban et~al.}(2019)\citenamefont{Esteban,
  Gonzalez-Garcia, Hernandez-Cabezudo, Maltoni, and Schwetz}}]{Esteban:2018azc}
\bibinfo{author}{\bibfnamefont{I.}~\bibnamefont{Esteban}},
  \bibinfo{author}{\bibfnamefont{M.~C.} \bibnamefont{Gonzalez-Garcia}},
  \bibinfo{author}{\bibfnamefont{A.}~\bibnamefont{Hernandez-Cabezudo}},
  \bibinfo{author}{\bibfnamefont{M.}~\bibnamefont{Maltoni}}, \bibnamefont{and}
  \bibinfo{author}{\bibfnamefont{T.}~\bibnamefont{Schwetz}},
  \bibinfo{journal}{JHEP} \textbf{\bibinfo{volume}{01}}, \bibinfo{pages}{106}
  (\bibinfo{year}{2019}), \eprint{1811.05487}.

\bibitem[{\citenamefont{de~Salas et~al.}(2021)\citenamefont{de~Salas, Forero,
  Gariazzo, Mart\'\i{}nez-Mirav\'e, Mena, Ternes, T\'ortola, and
  Valle}}]{deSalas:2020pgw}
\bibinfo{author}{\bibfnamefont{P.~F.} \bibnamefont{de~Salas}},
  \bibinfo{author}{\bibfnamefont{D.~V.} \bibnamefont{Forero}},
  \bibinfo{author}{\bibfnamefont{S.}~\bibnamefont{Gariazzo}},
  \bibinfo{author}{\bibfnamefont{P.}~\bibnamefont{Mart\'\i{}nez-Mirav\'e}},
  \bibinfo{author}{\bibfnamefont{O.}~\bibnamefont{Mena}},
  \bibinfo{author}{\bibfnamefont{C.~A.} \bibnamefont{Ternes}},
  \bibinfo{author}{\bibfnamefont{M.}~\bibnamefont{T\'ortola}},
  \bibnamefont{and} \bibinfo{author}{\bibfnamefont{J.~W.~F.}
  \bibnamefont{Valle}}, \bibinfo{journal}{JHEP} \textbf{\bibinfo{volume}{02}},
  \bibinfo{pages}{071} (\bibinfo{year}{2021}), \eprint{2006.11237}.

\bibitem[{\citenamefont{Aghanim et~al.}(2018)}]{Aghanim:2018eyx}
\bibinfo{author}{\bibfnamefont{N.}~\bibnamefont{Aghanim}} \bibnamefont{et~al.}
  (\bibinfo{collaboration}{Planck}) (\bibinfo{year}{2018}),
  \eprint{1807.06209}.

\bibitem[{\citenamefont{Brodsky and De~Rafael}(1968)}]{Brodsky:1967sr}
\bibinfo{author}{\bibfnamefont{S.~J.} \bibnamefont{Brodsky}} \bibnamefont{and}
  \bibinfo{author}{\bibfnamefont{E.}~\bibnamefont{De~Rafael}},
  \bibinfo{journal}{Phys. Rev.} \textbf{\bibinfo{volume}{168}},
  \bibinfo{pages}{1620} (\bibinfo{year}{1968}).

\bibitem[{\citenamefont{Baek and Ko}(2009)}]{Baek:2008nz}
\bibinfo{author}{\bibfnamefont{S.}~\bibnamefont{Baek}} \bibnamefont{and}
  \bibinfo{author}{\bibfnamefont{P.}~\bibnamefont{Ko}}, \bibinfo{journal}{JCAP}
  \textbf{\bibinfo{volume}{10}}, \bibinfo{pages}{011} (\bibinfo{year}{2009}),
  \eprint{0811.1646}.

\bibitem[{\citenamefont{Calibbi et~al.}(2018)\citenamefont{Calibbi, Ziegler,
  and Zupan}}]{Calibbi:2018rzv}
\bibinfo{author}{\bibfnamefont{L.}~\bibnamefont{Calibbi}},
  \bibinfo{author}{\bibfnamefont{R.}~\bibnamefont{Ziegler}}, \bibnamefont{and}
  \bibinfo{author}{\bibfnamefont{J.}~\bibnamefont{Zupan}},
  \bibinfo{journal}{JHEP} \textbf{\bibinfo{volume}{07}}, \bibinfo{pages}{046}
  (\bibinfo{year}{2018}), \eprint{1804.00009}.

\bibitem[{\citenamefont{Plumacher}(1997)}]{Plumacher:1996kc}
\bibinfo{author}{\bibfnamefont{M.}~\bibnamefont{Plumacher}},
  \bibinfo{journal}{Z. Phys. C} \textbf{\bibinfo{volume}{74}},
  \bibinfo{pages}{549} (\bibinfo{year}{1997}), \eprint{hep-ph/9604229}.

\bibitem[{\citenamefont{Buchmuller et~al.}(2002)\citenamefont{Buchmuller,
  Di~Bari, and Plumacher}}]{Buchmuller:2002rq}
\bibinfo{author}{\bibfnamefont{W.}~\bibnamefont{Buchmuller}},
  \bibinfo{author}{\bibfnamefont{P.}~\bibnamefont{Di~Bari}}, \bibnamefont{and}
  \bibinfo{author}{\bibfnamefont{M.}~\bibnamefont{Plumacher}},
  \bibinfo{journal}{Nucl. Phys.} \textbf{\bibinfo{volume}{B643}},
  \bibinfo{pages}{367} (\bibinfo{year}{2002}), \bibinfo{note}{[Erratum: Nucl.
  Phys.B793,362(2008)]}, \eprint{hep-ph/0205349}.

\bibitem[{\citenamefont{Pilaftsis}(1997)}]{Pilaftsis:1997jf}
\bibinfo{author}{\bibfnamefont{A.}~\bibnamefont{Pilaftsis}},
  \bibinfo{journal}{Phys. Rev. D} \textbf{\bibinfo{volume}{56}},
  \bibinfo{pages}{5431} (\bibinfo{year}{1997}), \eprint{hep-ph/9707235}.

\bibitem[{\citenamefont{Heeck and Teresi}(2016)}]{Heeck:2016oda}
\bibinfo{author}{\bibfnamefont{J.}~\bibnamefont{Heeck}} \bibnamefont{and}
  \bibinfo{author}{\bibfnamefont{D.}~\bibnamefont{Teresi}},
  \bibinfo{journal}{Phys. Rev. D} \textbf{\bibinfo{volume}{94}},
  \bibinfo{pages}{095024} (\bibinfo{year}{2016}), \eprint{1609.03594}.

\bibitem[{\citenamefont{Iso et~al.}(2011)\citenamefont{Iso, Okada, and
  Orikasa}}]{Iso:2010mv}
\bibinfo{author}{\bibfnamefont{S.}~\bibnamefont{Iso}},
  \bibinfo{author}{\bibfnamefont{N.}~\bibnamefont{Okada}}, \bibnamefont{and}
  \bibinfo{author}{\bibfnamefont{Y.}~\bibnamefont{Orikasa}},
  \bibinfo{journal}{Phys. Rev. D} \textbf{\bibinfo{volume}{83}},
  \bibinfo{pages}{093011} (\bibinfo{year}{2011}), \eprint{1011.4769}.

\bibitem[{\citenamefont{Okada et~al.}(2012)\citenamefont{Okada, Orikasa, and
  Yamada}}]{Okada:2012fs}
\bibinfo{author}{\bibfnamefont{N.}~\bibnamefont{Okada}},
  \bibinfo{author}{\bibfnamefont{Y.}~\bibnamefont{Orikasa}}, \bibnamefont{and}
  \bibinfo{author}{\bibfnamefont{T.}~\bibnamefont{Yamada}},
  \bibinfo{journal}{Phys. Rev. D} \textbf{\bibinfo{volume}{86}},
  \bibinfo{pages}{076003} (\bibinfo{year}{2012}), \eprint{1207.1510}.

\bibitem[{\citenamefont{Dev et~al.}(2018{\natexlab{b}})\citenamefont{Dev,
  Mohapatra, and Zhang}}]{Dev:2017xry}
\bibinfo{author}{\bibfnamefont{P.~S.~B.} \bibnamefont{Dev}},
  \bibinfo{author}{\bibfnamefont{R.~N.} \bibnamefont{Mohapatra}},
  \bibnamefont{and} \bibinfo{author}{\bibfnamefont{Y.}~\bibnamefont{Zhang}},
  \bibinfo{journal}{JHEP} \textbf{\bibinfo{volume}{03}}, \bibinfo{pages}{122}
  (\bibinfo{year}{2018}{\natexlab{b}}), \eprint{1711.07634}.

\bibitem[{\citenamefont{Mahanta and Borah}(2021)}]{Mahanta:2021plx}
\bibinfo{author}{\bibfnamefont{D.}~\bibnamefont{Mahanta}} \bibnamefont{and}
  \bibinfo{author}{\bibfnamefont{D.}~\bibnamefont{Borah}}
  (\bibinfo{year}{2021}), \eprint{2101.02092}.

\bibitem[{\citenamefont{Fileviez~P\'erez
  et~al.}(2021)\citenamefont{Fileviez~P\'erez, Murgui, and
  Plascencia}}]{FileviezPerez:2021ycy}
\bibinfo{author}{\bibfnamefont{P.}~\bibnamefont{Fileviez~P\'erez}},
  \bibinfo{author}{\bibfnamefont{C.}~\bibnamefont{Murgui}}, \bibnamefont{and}
  \bibinfo{author}{\bibfnamefont{A.~D.} \bibnamefont{Plascencia}}
  (\bibinfo{year}{2021}), \eprint{2103.13397}.

\bibitem[{\citenamefont{Altmannshofer et~al.}(2014)\citenamefont{Altmannshofer,
  Gori, Pospelov, and Yavin}}]{Altmannshofer:2014pba}
\bibinfo{author}{\bibfnamefont{W.}~\bibnamefont{Altmannshofer}},
  \bibinfo{author}{\bibfnamefont{S.}~\bibnamefont{Gori}},
  \bibinfo{author}{\bibfnamefont{M.}~\bibnamefont{Pospelov}}, \bibnamefont{and}
  \bibinfo{author}{\bibfnamefont{I.}~\bibnamefont{Yavin}},
  \bibinfo{journal}{Phys. Rev. Lett.} \textbf{\bibinfo{volume}{113}},
  \bibinfo{pages}{091801} (\bibinfo{year}{2014}), \eprint{1406.2332}.

\bibitem[{\citenamefont{Sirunyan et~al.}(2019)}]{CMS:2018gao}
\bibinfo{author}{\bibfnamefont{A.~M.} \bibnamefont{Sirunyan}}
  \bibnamefont{et~al.} (\bibinfo{collaboration}{CMS}), \bibinfo{journal}{Phys.
  Lett. B} \textbf{\bibinfo{volume}{792}}, \bibinfo{pages}{345}
  (\bibinfo{year}{2019}), \eprint{1808.03684}.

\bibitem[{\citenamefont{Amhis et~al.}(2021)}]{HFLAV:2019otj}
\bibinfo{author}{\bibfnamefont{Y.~S.} \bibnamefont{Amhis}} \bibnamefont{et~al.}
  (\bibinfo{collaboration}{HFLAV}), \bibinfo{journal}{Eur. Phys. J. C}
  \textbf{\bibinfo{volume}{81}}, \bibinfo{pages}{226} (\bibinfo{year}{2021}),
  \eprint{1909.12524}.

\bibitem[{\citenamefont{Robens and Stefaniak}(2015)}]{Robens:2015gla}
\bibinfo{author}{\bibfnamefont{T.}~\bibnamefont{Robens}} \bibnamefont{and}
  \bibinfo{author}{\bibfnamefont{T.}~\bibnamefont{Stefaniak}},
  \bibinfo{journal}{Eur. Phys. J.} \textbf{\bibinfo{volume}{C75}},
  \bibinfo{pages}{104} (\bibinfo{year}{2015}), \eprint{1501.02234}.

\bibitem[{\citenamefont{Chalons et~al.}(2016)\citenamefont{Chalons, Lopez-Val,
  Robens, and Stefaniak}}]{Chalons:2016jeu}
\bibinfo{author}{\bibfnamefont{G.}~\bibnamefont{Chalons}},
  \bibinfo{author}{\bibfnamefont{D.}~\bibnamefont{Lopez-Val}},
  \bibinfo{author}{\bibfnamefont{T.}~\bibnamefont{Robens}}, \bibnamefont{and}
  \bibinfo{author}{\bibfnamefont{T.}~\bibnamefont{Stefaniak}},
  \bibinfo{journal}{PoS} \textbf{\bibinfo{volume}{ICHEP2016}},
  \bibinfo{pages}{1180} (\bibinfo{year}{2016}), \eprint{1611.03007}.

\bibitem[{\citenamefont{López-Val and Robens}(2014)}]{Lopez-Val:2014jva}
\bibinfo{author}{\bibfnamefont{D.}~\bibnamefont{López-Val}} \bibnamefont{and}
  \bibinfo{author}{\bibfnamefont{T.}~\bibnamefont{Robens}},
  \bibinfo{journal}{Phys. Rev. D} \textbf{\bibinfo{volume}{90}},
  \bibinfo{pages}{114018} (\bibinfo{year}{2014}), \eprint{1406.1043}.

\bibitem[{\citenamefont{Khachatryan et~al.}(2015)}]{Khachatryan:2015cwa}
\bibinfo{author}{\bibfnamefont{V.}~\bibnamefont{Khachatryan}}
  \bibnamefont{et~al.} (\bibinfo{collaboration}{CMS}), \bibinfo{journal}{JHEP}
  \textbf{\bibinfo{volume}{10}}, \bibinfo{pages}{144} (\bibinfo{year}{2015}),
  \eprint{1504.00936}.

\bibitem[{\citenamefont{Strassler and Zurek}(2008)}]{Strassler:2006ri}
\bibinfo{author}{\bibfnamefont{M.~J.} \bibnamefont{Strassler}}
  \bibnamefont{and} \bibinfo{author}{\bibfnamefont{K.~M.} \bibnamefont{Zurek}},
  \bibinfo{journal}{Phys. Lett.} \textbf{\bibinfo{volume}{B661}},
  \bibinfo{pages}{263} (\bibinfo{year}{2008}), \eprint{hep-ph/0605193}.

\bibitem[{\citenamefont{M\"uhlleitner et~al.}(2017)\citenamefont{M\"uhlleitner,
  Sampaio, Santos, and Wittbrodt}}]{Muhlleitner:2017dkd}
\bibinfo{author}{\bibfnamefont{M.}~\bibnamefont{M\"uhlleitner}},
  \bibinfo{author}{\bibfnamefont{M.~O.~P.} \bibnamefont{Sampaio}},
  \bibinfo{author}{\bibfnamefont{R.}~\bibnamefont{Santos}}, \bibnamefont{and}
  \bibinfo{author}{\bibfnamefont{J.}~\bibnamefont{Wittbrodt}},
  \bibinfo{journal}{JHEP} \textbf{\bibinfo{volume}{08}}, \bibinfo{pages}{132}
  (\bibinfo{year}{2017}), \eprint{1703.07750}.

\bibitem[{\citenamefont{Haller et~al.}(2018)\citenamefont{Haller, Hoecker,
  Kogler, M\"onig, Peiffer, and Stelzer}}]{Haller:2018nnx}
\bibinfo{author}{\bibfnamefont{J.}~\bibnamefont{Haller}},
  \bibinfo{author}{\bibfnamefont{A.}~\bibnamefont{Hoecker}},
  \bibinfo{author}{\bibfnamefont{R.}~\bibnamefont{Kogler}},
  \bibinfo{author}{\bibfnamefont{K.}~\bibnamefont{M\"onig}},
  \bibinfo{author}{\bibfnamefont{T.}~\bibnamefont{Peiffer}}, \bibnamefont{and}
  \bibinfo{author}{\bibfnamefont{J.}~\bibnamefont{Stelzer}},
  \bibinfo{journal}{Eur. Phys. J. C} \textbf{\bibinfo{volume}{78}},
  \bibinfo{pages}{675} (\bibinfo{year}{2018}), \eprint{1803.01853}.

\bibitem[{\citenamefont{Misiak and Steinhauser}(2017)}]{Misiak:2017bgg}
\bibinfo{author}{\bibfnamefont{M.}~\bibnamefont{Misiak}} \bibnamefont{and}
  \bibinfo{author}{\bibfnamefont{M.}~\bibnamefont{Steinhauser}},
  \bibinfo{journal}{Eur. Phys. J. C} \textbf{\bibinfo{volume}{77}},
  \bibinfo{pages}{201} (\bibinfo{year}{2017}), \eprint{1702.04571}.

\bibitem[{\citenamefont{Arhrib et~al.}(2018)\citenamefont{Arhrib, Benbrik,
  Moretti, Rouchad, Yan, and Zhang}}]{Arhrib:2017uon}
\bibinfo{author}{\bibfnamefont{A.}~\bibnamefont{Arhrib}},
  \bibinfo{author}{\bibfnamefont{R.}~\bibnamefont{Benbrik}},
  \bibinfo{author}{\bibfnamefont{S.}~\bibnamefont{Moretti}},
  \bibinfo{author}{\bibfnamefont{A.}~\bibnamefont{Rouchad}},
  \bibinfo{author}{\bibfnamefont{Q.-S.} \bibnamefont{Yan}}, \bibnamefont{and}
  \bibinfo{author}{\bibfnamefont{X.}~\bibnamefont{Zhang}},
  \bibinfo{journal}{JHEP} \textbf{\bibinfo{volume}{07}}, \bibinfo{pages}{007}
  (\bibinfo{year}{2018}), \eprint{1712.05332}.

\bibitem[{\citenamefont{Lees et~al.}(2016)}]{TheBABAR:2016rlg}
\bibinfo{author}{\bibfnamefont{J.}~\bibnamefont{Lees}} \bibnamefont{et~al.}
  (\bibinfo{collaboration}{BaBar}), \bibinfo{journal}{Phys. Rev. D}
  \textbf{\bibinfo{volume}{94}}, \bibinfo{pages}{011102}
  (\bibinfo{year}{2016}), \eprint{1606.03501}.

\bibitem[{\citenamefont{Bauer et~al.}(2020)\citenamefont{Bauer, Foldenauer, and
  Jaeckel}}]{Bauer:2018onh}
\bibinfo{author}{\bibfnamefont{M.}~\bibnamefont{Bauer}},
  \bibinfo{author}{\bibfnamefont{P.}~\bibnamefont{Foldenauer}},
  \bibnamefont{and} \bibinfo{author}{\bibfnamefont{J.}~\bibnamefont{Jaeckel}},
  \bibinfo{journal}{JHEP} \textbf{\bibinfo{volume}{18}}, \bibinfo{pages}{094}
  (\bibinfo{year}{2020}), \eprint{1803.05466}.

\bibitem[{\citenamefont{Kamada et~al.}(2018)\citenamefont{Kamada, Kaneta,
  Yanagi, and Yu}}]{Kamada:2018zxi}
\bibinfo{author}{\bibfnamefont{A.}~\bibnamefont{Kamada}},
  \bibinfo{author}{\bibfnamefont{K.}~\bibnamefont{Kaneta}},
  \bibinfo{author}{\bibfnamefont{K.}~\bibnamefont{Yanagi}}, \bibnamefont{and}
  \bibinfo{author}{\bibfnamefont{H.-B.} \bibnamefont{Yu}},
  \bibinfo{journal}{JHEP} \textbf{\bibinfo{volume}{06}}, \bibinfo{pages}{117}
  (\bibinfo{year}{2018}), \eprint{1805.00651}.

\bibitem[{\citenamefont{Ibe et~al.}(2020)\citenamefont{Ibe, Kobayashi,
  Nakayama, and Shirai}}]{Ibe:2019gpv}
\bibinfo{author}{\bibfnamefont{M.}~\bibnamefont{Ibe}},
  \bibinfo{author}{\bibfnamefont{S.}~\bibnamefont{Kobayashi}},
  \bibinfo{author}{\bibfnamefont{Y.}~\bibnamefont{Nakayama}}, \bibnamefont{and}
  \bibinfo{author}{\bibfnamefont{S.}~\bibnamefont{Shirai}},
  \bibinfo{journal}{JHEP} \textbf{\bibinfo{volume}{04}}, \bibinfo{pages}{009}
  (\bibinfo{year}{2020}), \eprint{1912.12152}.

\bibitem[{\citenamefont{Escudero et~al.}(2019)\citenamefont{Escudero, Hooper,
  Krnjaic, and Pierre}}]{Escudero:2019gzq}
\bibinfo{author}{\bibfnamefont{M.}~\bibnamefont{Escudero}},
  \bibinfo{author}{\bibfnamefont{D.}~\bibnamefont{Hooper}},
  \bibinfo{author}{\bibfnamefont{G.}~\bibnamefont{Krnjaic}}, \bibnamefont{and}
  \bibinfo{author}{\bibfnamefont{M.}~\bibnamefont{Pierre}},
  \bibinfo{journal}{JHEP} \textbf{\bibinfo{volume}{03}}, \bibinfo{pages}{071}
  (\bibinfo{year}{2019}), \eprint{1901.02010}.

\bibitem[{\citenamefont{Akimov et~al.}(2017)}]{Akimov:2017ade}
\bibinfo{author}{\bibfnamefont{D.}~\bibnamefont{Akimov}} \bibnamefont{et~al.}
  (\bibinfo{collaboration}{COHERENT}), \bibinfo{journal}{Science}
  \textbf{\bibinfo{volume}{357}}, \bibinfo{pages}{1123} (\bibinfo{year}{2017}),
  \eprint{1708.01294}.

\bibitem[{\citenamefont{Akimov et~al.}(2021)}]{Akimov:2020pdx}
\bibinfo{author}{\bibfnamefont{D.}~\bibnamefont{Akimov}} \bibnamefont{et~al.}
  (\bibinfo{collaboration}{COHERENT}), \bibinfo{journal}{Phys. Rev. Lett.}
  \textbf{\bibinfo{volume}{126}}, \bibinfo{pages}{012002}
  (\bibinfo{year}{2021}), \eprint{2003.10630}.

\bibitem[{\citenamefont{Cadeddu et~al.}(2021)\citenamefont{Cadeddu, Cargioli,
  Dordei, Giunti, Li, Picciau, and Zhang}}]{Cadeddu:2020nbr}
\bibinfo{author}{\bibfnamefont{M.}~\bibnamefont{Cadeddu}},
  \bibinfo{author}{\bibfnamefont{N.}~\bibnamefont{Cargioli}},
  \bibinfo{author}{\bibfnamefont{F.}~\bibnamefont{Dordei}},
  \bibinfo{author}{\bibfnamefont{C.}~\bibnamefont{Giunti}},
  \bibinfo{author}{\bibfnamefont{Y.~F.} \bibnamefont{Li}},
  \bibinfo{author}{\bibfnamefont{E.}~\bibnamefont{Picciau}}, \bibnamefont{and}
  \bibinfo{author}{\bibfnamefont{Y.~Y.} \bibnamefont{Zhang}},
  \bibinfo{journal}{JHEP} \textbf{\bibinfo{volume}{01}}, \bibinfo{pages}{116}
  (\bibinfo{year}{2021}), \eprint{2008.05022}.

\bibitem[{\citenamefont{Banerjee et~al.}(2021)\citenamefont{Banerjee, Dutta,
  and Roy}}]{Banerjee:2021laz}
\bibinfo{author}{\bibfnamefont{H.}~\bibnamefont{Banerjee}},
  \bibinfo{author}{\bibfnamefont{B.}~\bibnamefont{Dutta}}, \bibnamefont{and}
  \bibinfo{author}{\bibfnamefont{S.}~\bibnamefont{Roy}} (\bibinfo{year}{2021}),
  \eprint{2103.10196}.

\bibitem[{\citenamefont{Krnjaic et~al.}(2020)\citenamefont{Krnjaic,
  Marques-Tavares, Redigolo, and Tobioka}}]{Krnjaic:2019rsv}
\bibinfo{author}{\bibfnamefont{G.}~\bibnamefont{Krnjaic}},
  \bibinfo{author}{\bibfnamefont{G.}~\bibnamefont{Marques-Tavares}},
  \bibinfo{author}{\bibfnamefont{D.}~\bibnamefont{Redigolo}}, \bibnamefont{and}
  \bibinfo{author}{\bibfnamefont{K.}~\bibnamefont{Tobioka}},
  \bibinfo{journal}{Phys. Rev. Lett.} \textbf{\bibinfo{volume}{124}},
  \bibinfo{pages}{041802} (\bibinfo{year}{2020}), \eprint{1902.07715}.

\bibitem[{\citenamefont{Gninenko et~al.}(2015)\citenamefont{Gninenko,
  Krasnikov, and Matveev}}]{Gninenko:2014pea}
\bibinfo{author}{\bibfnamefont{S.}~\bibnamefont{Gninenko}},
  \bibinfo{author}{\bibfnamefont{N.}~\bibnamefont{Krasnikov}},
  \bibnamefont{and} \bibinfo{author}{\bibfnamefont{V.}~\bibnamefont{Matveev}},
  \bibinfo{journal}{Phys. Rev. D} \textbf{\bibinfo{volume}{91}},
  \bibinfo{pages}{095015} (\bibinfo{year}{2015}), \eprint{1412.1400}.

\bibitem[{\citenamefont{Gninenko and Krasnikov}(2018)}]{Gninenko:2018tlp}
\bibinfo{author}{\bibfnamefont{S.}~\bibnamefont{Gninenko}} \bibnamefont{and}
  \bibinfo{author}{\bibfnamefont{N.}~\bibnamefont{Krasnikov}},
  \bibinfo{journal}{Phys. Lett. B} \textbf{\bibinfo{volume}{783}},
  \bibinfo{pages}{24} (\bibinfo{year}{2018}), \eprint{1801.10448}.

\bibitem[{\citenamefont{Branco et~al.}(2012)\citenamefont{Branco, Ferreira,
  Lavoura, Rebelo, Sher, and Silva}}]{Branco:2011iw}
\bibinfo{author}{\bibfnamefont{G.~C.} \bibnamefont{Branco}},
  \bibinfo{author}{\bibfnamefont{P.~M.} \bibnamefont{Ferreira}},
  \bibinfo{author}{\bibfnamefont{L.}~\bibnamefont{Lavoura}},
  \bibinfo{author}{\bibfnamefont{M.~N.} \bibnamefont{Rebelo}},
  \bibinfo{author}{\bibfnamefont{M.}~\bibnamefont{Sher}}, \bibnamefont{and}
  \bibinfo{author}{\bibfnamefont{J.~P.} \bibnamefont{Silva}},
  \bibinfo{journal}{Phys. Rept.} \textbf{\bibinfo{volume}{516}},
  \bibinfo{pages}{1} (\bibinfo{year}{2012}), \eprint{1106.0034}.

\end{thebibliography}

\end{document}